\documentclass[11pt,a4paper]{article}

\usepackage{comment}
\usepackage{graphicx}
\usepackage{amsmath}
\usepackage{amsfonts}
\usepackage{amssymb}
\usepackage{enumerate}
\usepackage{lscape}
\usepackage{longtable}
\usepackage{rotating}
\usepackage{color}
\usepackage{url}
\usepackage{rotating}
\usepackage{algpseudocode}
\usepackage[ruled]{algorithm}
\usepackage{pgf,tikz}
\usepackage{subfig}
\usepackage[titletoc]{appendix}
\usepackage{wrapfig}
\usepackage{adjustbox}
\usepackage{mathrsfs}
\usepackage{pgfplots}
\usetikzlibrary{arrows}
\usepackage{color}

\usepackage{bbm}

\usepackage[tableposition=below]{caption}
\usepackage{booktabs}
\usepackage{multirow}
\usepackage{mathtools}
\usepackage{makecell}

\usepackage[sort,numbers]{natbib}
\usepackage{eqparbox}%

\numberwithin{equation}{section}%

\usepackage[symbol]{footmisc}%

\DeclareMathOperator{\UAS}{UAS}
\DeclareMathOperator{\score}{score}

\newcommand{\beqn}[1]{\begin{equation}\label{#1}}
\newcommand{\eeqn}{\end{equation}}

\definecolor{darkgreen}{rgb}{0,0.6,0}
\definecolor{aau2}{rgb}{0.0, 0.5, 0.69}
\definecolor{aau3}{rgb}{0.0, 0.53, 0.74}
\definecolor{aau4}{rgb}{0.0, 0.48, 0.65}
\definecolor{aau5}{rgb}{0.0, 0.45, 0.73}
\definecolor{rsap}{RGB}{130, 36, 51}
\definecolor{gsap}{RGB}{112, 164, 137}
\definecolor{tud}{rgb}{0.43,0.73,0.11}
\definecolor{verde}{rgb}{0.33,0.53,0.11}

\definecolor{ttffqq}{rgb}{0.0, 0.48, 0.65} %{rgb}{0.43,0.73,0.11}
\definecolor{ffqqqq}{rgb}{0.0, 0.5, 0.69} %{rgb}{1,0,0}
\usetikzlibrary{arrows}

\tikzstyle{decision} = [diamond, draw, fill=blue!20,
text width=4.5em, text badly centered, node distance=3cm, inner sep=0pt]
\tikzstyle{block} = [rectangle, draw, fill=blue!20,
text centered, rounded corners, minimum height=4em]
\tikzstyle{line} = [draw, -latex']
\tikzstyle{cloud} = [draw, ellipse,fill=red!20, node distance=3cm,
minimum height=2em]
\tikzstyle{cloud2} = [draw, ellipse,fill=green!20, node distance=3cm,
minimum height=2em]

\begin{document}
	
	\title{Why is soccer so popular: \\ Understanding underdog achievement and randomness \\ in team ball sports}
 
	% \author{		
 %            L. N. Vicente\thanks{Department of Industrial and Systems Engineering, Lehigh University, Bethlehem, PA 18015-1582, USA ({\tt lnv@lehigh.edu}).}
	% 	\and
	% 	T. N. Alleck\thanks{Department of Industrial and Systems Engineering, Lehigh University, Bethlehem, PA 18015-1582, USA ({\tt tna324@lehigh.edu}).}
	% 	\and
 %  		T. Giovannelli\thanks{Department of Industrial and Systems Engineering, Lehigh University, Bethlehem, PA 18015-1582, USA ({\tt tog220@lehigh.edu}).}
 %            \and
 %  		R. Mitchell\thanks{Department of Industrial and Systems Engineering, Lehigh University, Bethlehem, PA 18015-1582, USA ({\tt rgm424@lehigh.edu}).}            
 %            \and
 %            O. Remen\thanks{Department of Industrial and Systems Engineering, Lehigh University, Bethlehem, PA 18015-1582, USA  ({\tt orr224@lehigh.edu}).}
 %        }

\author{
    L. N. Vicente \and
    T. N. Alleck \and
    T. Giovannelli \and
    R. Mitchell \and
    O. Remen
}

	\maketitle
    
\newcommand{\affiliation}{Department of Industrial and Systems Engineering, Lehigh University, Bethlehem, PA 18015-1582, USA}

\footnotetext{\affiliation}
\footnotetext{Emails: lnv@lehigh.edu, tna324@lehigh.edu, tog220@lehigh.edu, rgm424@lehigh.edu, orr224@lehigh.edu}
	
	\begin{abstract}
        	In this paper, we examine team ball sports to investigate how the likelihood of weaker teams winning against stronger ones, referred to as underdog achievement, is influenced by inherent randomness factors that affect match outcomes in such sports. To address our research question, we collected data on match scores and computed corresponding team rankings from major international competitions for~12 popular team ball sports: basketball, cricket, field hockey, futsal, handball, ice hockey, lacrosse, roller hockey, rugby, soccer, volleyball, and water polo. Then, we developed an underdog achievement score to identify the sports with the highest occurrences of weaker teams prevailing over stronger ones, and we designed a randomness model consisting of factors that contribute to unexpected match outcomes within each sport. 
            Our findings indicate that soccer is among the sports in which a weaker team is most likely to win. Through principal component analysis~(PCA) and correlation analysis, we demonstrate that our randomness model can explain such a phenomenon, showing that the underdog achievement can be attributed to numerous factors that can randomly influence match outcomes.
	\end{abstract}

%%%%%%%%%%%%%%%%%%%%%%%%%%%%%%%%%%%%%%%%%%%%%%%
\section{Introduction} 
%%%%%%%%%%%%%%%%%%%%%%%%%%%%%%%%%%%%%%%%%%%%%%%

Team ball sports have been the subject of growing research over the past decade, and most of the papers are within the domain of sports medicine literature~\cite{HSarmento_etal_2022}. In our work, we do not focus on medicine applications. Instead, we aim to provide novel insights into team ball sports by determining which sports are more likely to see weaker teams win, referred to as underdog achievement, and which randomness factors contribute to such an outcome. 
% with the goal of investigating any potential relationship between underdog wins and randomness. 
Since certain teams consistently outperform others in all sports, it is clear that outcomes of matches are not purely random. The question of whether a team wins due to chance or their own skills has been discussed in non-scholarly books. For example, in~\cite{DSally_CAnderson_2013}, the authors claim that soccer, which is universally recognized as the world's most popular sport~\cite{Dvorak_etal_2004}, is also the most random, and its inherent randomness is what makes soccer so popular. 
Similar conclusions were confirmed in the academic literature by~\cite{EBen-Naim_etal_2006}, where the authors analyzed the English Football Association and four major North American professional sports leagues (MLB for baseball, NBA for basketball, NFL for American football, and~NHL for hockey), finding that soccer is the sport with the most random outcomes.
In general, the randomness element adds excitement and unpredictability, making a sport enjoyable to watch. However, one should also notice that excessive randomness can diminish interest, as viewers prefer a balance between unpredictability and skill~\cite{MMauboussin_2013}.

A novel research question is whether a weak team is more likely to win than a strong one as a consequence of certain random and situational conditions. Such a question has recently emerged as a research focus and has been studied by~\cite{FWunderlich_etal_2021}, where the authors use data from the English Premier League to show that the influence of randomness on goals in soccer decreases as the match progresses. Such a decreasing trend was observed to be disadvantageous for weaker teams, as they rely more on randomness to score. The study in~\cite{FWunderlich_etal_2021} also identifies variables of randomness that affect the outcome during a match and cannot be entirely attributed to skills. Examples of such variables include the degree of involvement of the defending team and the chances to score goals from outside the penalty area. Additionally, the analysis includes situational variables that may influence the outcome by affecting the motivation of players, such as match location and current score. We conclude our literature review with~\cite{MJLopez_etal_2017}, where a Bayesian state-space model was proposed to study the randomness in match outcomes for four major North American professional sports leagues
(once again, MLB, NBA, NFL, and~NHL). Probability-based metrics derived from betting market data were used to quantify the influence of chance on the outcomes. Their findings indicate that the~MLB and~NHL exhibit the highest levels of randomness in match outcomes (and we found a similar result for ice hockey in our study). However, soccer was not included in their analysis.

Previous studies examining underdog achievement have focused on a limited range of team ball sports and have not systematically identified the randomness factors contributing to underdog victories. The ultimate goal of our paper is to investigate how the likelihood of weaker teams winning against stronger ones is influenced by inherent randomness factors in team ball sports. To achieve our goal, we selected major international competitions for~12 popular team ball sports: basketball, cricket, field hockey, futsal, handball, ice hockey\footnote{We include ice hockey among the team ball sports, even though it technically uses a puck.}, lacrosse, roller hockey, rugby, soccer, volleyball, and water polo. The official names of the competitions selected for each sport are included in Table~\ref{tab:competitions}. Note that such competitions are all men's events, for which more data is available. We notice that certain popular team ball sports, such as American football, baseball, and tennis, were not included in our analysis. This decision was made due to either the absence of international competitions for these sports or the smaller size of their teams compared to the sports considered in this paper (for instance, in tennis, teams consist of at most two players, whereas all the other sports discussed in our paper involve teams with much more than two players).

Our main contributions can be summarized as follows:
\begin{itemize}
    \item We collected a vast amount of data containing information related to match scores, and we computed corresponding team rankings for each edition of the competitions in Table~\ref{tab:competitions}. This data represents valuable information for researchers in the field of sports analytics.
    \item We developed an underdog achievement score to determine the sports with the highest and lowest occurrences of weaker teams defeating stronger ones when focusing on a much broader range of team ball sports than the ones considered in the literature. In accordance with the limited existing literature (again, see~\cite{DSally_CAnderson_2013,EBen-Naim_etal_2006}), soccer is among the sports with the highest underdog achievement. 
    \item We designed a randomness model consisting of~14 factors that contribute to unexpected match outcomes within each sport, providing quantitative values for each of the factors.
    \item We performed principal component analysis~(PCA) and correlation analysis to identify the randomness factors with the greatest impact on underdog achievement and demonstrate that our randomness model can explain underdog achievement.
    % Factors related to ball properties are responsible for the most variability among all the factors. 
    % The underdog achievement score was observed to be positively correlated with .  
\end{itemize}

Our paper is organized as follows. In Section~\ref{sec:data_collection}, we detail our data collection process. In Section~\ref{sec:underdog_achievement}, we develop an underdog achievement score and in Section~\ref{sec:randomness_model}, we present our randomness model. In Section~\ref{sec:PCA_analysis}, we perform~PCA and correlation analysis to demonstrate how our randomness model can explain underdog achievement. In Section~\ref{sec:conclusions}, we conclude our paper with some remarks and ideas for future work.

\begin{table}[h]
\centering
\begin{tabular}{|l|c|}
\hline
\textbf{Sport} & \textbf{Competition} \\
\hline
Basketball & Summer Olympic Games \\
Cricket & ICC Men's Cricket World Cup \\
Field Hockey & Men's FIH Hockey World Cup \\
Futsal & FIFA Futsal World Cup \\
Handball & Summer Olympic Games \\
Ice Hockey & Winter Olympic Games \\
Lacrosse & World Lacrosse Men's World Cup \\
Roller Hockey & World Skate Roller Hockey World Cup \\
Rugby & Rugby World Cup \\
Soccer & FIFA World Cup \\
Volleyball & FIVB Volleyball Men's World Cup \\
Water Polo & FINA Men's Water Polo World Cup \\
\hline
\end{tabular}
\caption{Major international competitions selected for the team ball sports included in our paper.}
\label{tab:competitions}
\end{table}

%%%%%%%%%%%%%%%%%%%%%%%%%%%%%%%%%%%%%%%%%%%%%%%
\section{Data collection: Match scores and team rankings}\label{sec:data_collection} 
%%%%%%%%%%%%%%%%%%%%%%%%%%%%%%%%%%%%%%%%%%%%%%%
To perform our analysis, for each sport, we collected real data on match scores and computed corresponding team rankings from all available editions of the major international sports competitions in Table~\ref{tab:competitions}. The complete table with the years of the editions of each competition is provided in Table~\ref{tab:competitions_app} of Appendix~\ref{sec:app}. 
% \textcolor{blue}{Each year corresponds to a different edition.} 
To obtain match score data for each edition, we conducted web scraping of match information from Wikipedia pages. 
% resulting in over~5600 matches. 
All this data was aggregated into a match score dataset, which contains information related to individual matches, including the names of the two opposing teams and their respective scores. 
Given the match score dataset, we then computed a team ranking for each edition. 
% For more details about the match score dataset and the team rankings, we refer to~[Our paper 2].
Finally, for each competition, we aggregated the team rankings across all the edition years included in Table~\ref{tab:competitions_app} into a weighted team ranking. Our code is publicly available on GitHub.\footnote{\url{https://github.com/thaksheel/randomness-team-ball-sports.git}}
% giving more weight to the latest editions.

We will now introduce some general notation that will allow us to formally describe the match score dataset, the team ranking for each edition, and the weighted team ranking. Denoting as~$\mathcal{S}$ the set of sports, let~$\mathcal{E}_s$ be the set of editions for the competition selected for sport~$s \in \mathcal{S}$ in Table~\ref{tab:competitions} (one can think of such a set as a set of edition years) and let~$\mathcal{P}_e^s = \{p_1,p_2,\ldots,p_{N_e^s}\}$ be the set of all teams playing in edition~$e \in \mathcal{E}_s$, where~$N_e^s$ is the total number of teams in that edition. 
% \textcolor{red}{The set of teams across all editions will be denoted as~$\mathcal{P} = \cup_{e \in \mathcal{E}_s} \mathcal{P}_e$.}
We will denote as~$\mathcal{M}_e^s \subseteq \mathcal{P}_e^s \times \mathcal{P}_e^s$ the set of matches in edition~$e$, represented as a set of tuples~$(i,j)$, where~$i$ and~$j$ are opposing teams belonging to~$\mathcal{P}_e^s$. For the rest of this section, we will omit the subscript~$s$ to simplify the notation. All the notation used in our paper is summarized in Table~\ref{tab:list_notation}.    

\begin{table}
    \small
    \centering
    \begin{tabular}{|l|l|} 
     \hline
     $\mathcal{S}$ & Set of sports.\\[1pt]
     \hline
     $\mathcal{E}$ & Set of editions for the given competition.\\[2pt]
     $\mathcal{P}_e = \{p_1,p_2,\ldots,p_{N_e}\}$ & Set of all teams playing in edition~$e \in \mathcal{E}$.\\[2pt]
     $\mathcal{M}_e \subseteq \mathcal{P}_e \times \mathcal{P}_e$ & Set of matches in edition~$e \in \mathcal{E}$.\\[2pt]
     $\score_{ij}^{e}(i), \, \score_{ij}^{e}(j)$ & Scores obtained by teams~$i$ and~$j$ when facing each other, \\[2pt] & where~$(i,j) \in \mathcal{M}_e$.\\[2pt]
     $\mathcal{D}_e$ &  Match score dataset for edition~$e \in \mathcal{E}$, as defined in~\eqref{eq:match_score_dataset}.\\[2pt]
     $\mathcal{R}_e$ & Team ranking for edition~$e \in \mathcal{E}$, as defined in~\eqref{eq:team_ranking}.\\[2pt]
     $\mathcal{E} = (e_1,e_2,\ldots,e_{|\mathcal{E}|})$ & Set of editions written as an ordered list of elements, where~$e_1$ is the \\[2pt] & earliest edition and~$e_{|\mathcal{E}|}$ is the most recent edition.\\[2pt]
     $\mathcal{P}_{\leqslant e}$ & Union of the sets of teams up to edition~$e \in \mathcal{E}$, as defined in~\eqref{eq:union_set_teams}.\\[2pt]
     $N_{\leqslant e}$ & Cardinality of the set~$\mathcal{P}_{\leqslant e}$.\\[2pt]
     $c(i,{\mathcal{R}_e})$ & Position of a team~$i \in \mathcal{P}_e$ in the team ranking~$\mathcal{R}_e$.\\[2pt]
     $\mathcal{W}_{\leqslant e}$ & Weighted team ranking up to edition~$e \in \mathcal{E}_s$, as defined in~\eqref{eq:weighted_team_ranking}.\\[2pt]
     $wr_{\leqslant e}$ & Sorting function used to assign each team to their position in~$\mathcal{W}_{\leqslant e}$,\\[2pt] & as defined in~\eqref{eq:weighted_ranking}.\\[2pt]
     $c(i,{\mathcal{W}_{\leqslant e}})$ & Position of a team~$i \in \mathcal{P}_{\leqslant e}$ in the weighted team ranking~$\mathcal{W}_{\leqslant e}$.\\[2pt]
     $|c(i,{\mathcal{R}_{e}}) - c(j,{\mathcal{R}_{e}})|$ & Rank difference between teams~$i \in \mathcal{P}_e$ and~$j \in \mathcal{P}_e$ in~$\mathcal{R}_{e}$.\\[2pt]
     $|c(i,{\mathcal{W}_{\leqslant e}}) - c(j,{\mathcal{W}_{\leqslant e}})|$ & Rank difference between teams~$i \in \mathcal{P}_{\leqslant e}$ and~$j \in  \mathcal{P}_{\leqslant e}$ in~$\mathcal{W}_{\leqslant e}$.\\[2pt]
     $\tau$ & Rank difference threshold used to identify weak teams, as defined in~\eqref{eq:suff_decrease}.\\[2pt]
     \hline
     $\lambda$ & Decay factor used in~$wr_{\leqslant e}$ to determine past edition relevance.\\
     \hline
    \end{tabular}
    \caption{Notation.}\label{tab:list_notation}
\end{table}

%%%%%%%%%%%%%%%%%%%% 
\paragraph{Match score dataset.}

For any edition~$e \in \mathcal{E}$ and match~$(i,j) \in \mathcal{M}_e$, let~$\score_{ij}^e(i)$ denote the score that team~$i$ obtained when playing against team~$j$ (for example, if~``5-6'' is the outcome of the match~$(i,j)$, then~$\score_{ij}^e(i) = 5$ and~$\score_{ij}^e(j) = 6$). The match score dataset for edition~$e$ can be represented by the following set
\begin{equation}\label{eq:match_score_dataset}
\mathcal{D}_e = \{(i,j,\score_{ij}^{e}(i),\score_{ij}^{e}(j)) ~|~ (i,j) \in \mathcal{M}_e\}.
\end{equation} 
To populate such a dataset, we web-scraped information related to individual matches from the Wikipedia pages corresponding to each edition year of a competition. All sports competitions typically include a group stage, where teams are divided into groups and each team plays against the others in its group to collect the maximum number of points and advance in the competition, and a knockout stage (or bracket stage), where teams are eliminated from the competition if they lose a match. The knockout stage typically consists of the following additional phases: rounds of~16, quarterfinals, semifinals, and finals.
For more details, we refer to~\cite{TNAlleck_etal_2024}.

%%%%%%%%%%%%%%%%%%%%
\paragraph{Team rankings.}

Based on the match score dataset~$\mathcal{D}_e$ in~\eqref{eq:match_score_dataset}, for each edition~$e \in \mathcal{E}$, we generated a team ranking by sorting teams based on the following criteria, listed in order of priority: number of matches played, number of victories, number of draws, number of losses, and total score across all matches. The sorting was in descending order for each criterion except for the number of losses, for which an ascending order was used. The choice to use the number of matches played as a sorting criterion is due to the fact that teams reaching the final stages of a competition typically play the largest number of matches, reflecting their strength. 

For each edition~$e$, the team ranking obtained through the sorting criteria above is represented as an ordered list of teams
\begin{equation}\label{eq:team_ranking}
\mathcal{R}_e = (i_1, i_2, \ldots, i_{N_e}),
\end{equation}
where~$i_j \in \mathcal{P}_e$ for any~$j \in \{1,2,\ldots,N_e\}$.

%%%%%%%%%%%%%%%%%%%%
\paragraph{Weighted team ranking.}

The weighted team ranking aggregates team rankings from the earliest available edition up to a given edition. When referring to the weighted team ranking, we will explicitly write the set of editions as an ordered list of elements as follows~$\mathcal{E} = (e_1,e_2,\ldots,e_{|\mathcal{E}|})$, where~$e_1$ is the earliest available edition and~$e_{|\mathcal{E}|}$ is the most recent edition. In such a case, for any~$h \in \{1,2,\ldots,|\mathcal{E}|\}$, we will denote as
\begin{equation}\label{eq:union_set_teams}
\mathcal{P}_{\leqslant e_h} = \cup \{\mathcal{P}_{e_{\bar{h}}} ~|~ e_{\bar{h}} \in \mathcal{E} \text{ and } \bar{h} \le h\}
\end{equation}
the union of the sets of teams from edition~$e_1$ to edition~$e_h$. We will denote as~$N_{\leqslant e_{h}}$ the cardinality of the set~$\mathcal{P}_{\leqslant e_{h}}$.

To build the weighted team ranking that aggregates team rankings up to an arbitrary edition~$e \in \mathcal{E}$, denoted as~$\mathcal{W}_{\leqslant e}$, we use a sorting function~$wr_{\leqslant e}: \mathcal{P}_{\leqslant e} \to \mathbb{R}$ to assign each team to their position in such a weighted team ranking. 
For any~$i \in \mathcal{P}_{\leqslant e}$, the higher the value of~$wr_{\leqslant e}(i)$, the higher the position of team~$i$ in the weighted team ranking. 
To show how we compute the weighted team ranking~$\mathcal{W}_{\leqslant e}$, let us denote as~$c(i,{\mathcal{R}_e}) \in [1,N_e]$ the position of a team~$i \in \mathcal{P}_e$ in the team ranking~$\mathcal{R}_e$ in~\eqref{eq:team_ranking}.
% where~$c_e(i) = 1$ indicates the top team in the ranking and~$c_e(i) = N_e$ refers to the last team. 
For any~$e_h \in \mathcal{E} = (e_1,e_2,\ldots,e_{|\mathcal{E}|})$, with~$h \in \{1,2,\ldots,|\mathcal{E}|\}$, we compute~$wr_{\leqslant e_h}(i)$ as follows
\begin{alignat}{2}\label{eq:weighted_ranking}
wr_{\leqslant e_h}(i) \;=\; 
\begin{cases}
(N_{e_h} - c(i,\mathcal{R}_{e_h})), \quad & \text{ if } i \in \mathcal{P}_{e_h} \text{ and } h = 1,\\
(N_{e_h} - c(i,\mathcal{R}_{e_{h}})) + \lambda \, wr_{\leqslant e_{h-1}}(i), \quad & \text{ if } i \in \mathcal{P}_{e_h} \text{ and } h \in \{2,\ldots,|\mathcal{E}|\},\\
0, \quad & \text{ if } i \not\in \mathcal{P}_{e_h},
\end{cases}
\end{alignat}
where~$\lambda \in [0,1]$ is a decay factor that dictates the rate at which past editions become irrelevant.

The weighted team ranking up to edition~$e$ is represented as an ordered list of teams 
\begin{equation}\label{eq:weighted_team_ranking}
\mathcal{W}_{\leqslant e} = (i_1, i_2, \ldots, i_{N_{\leqslant e}}),
\end{equation}
where $i_j \in \mathcal{P}_{\leqslant e}$ for any~$j \in \{1,2,\ldots,N_{\leqslant e}\}$ and~$wr_{\leqslant e}(i_j) \ge wr_{\leqslant e}(i_{j+1})$ for any~$e \in \mathcal{E}$ and~$j \in \{1,2,\ldots,N_{\leqslant e}-1\}$.

\section{Underdog achievement}\label{sec:underdog_achievement}
%%%%%%%%%%%%%%%%%%%%%%%%%%%%%%%%%%%%%%%%%%%%%%%

To quantify the underdog achievement, we first need to determine criteria that allow us to distinguish weak teams from strong ones. As noticed in the literature related to soccer~\cite{FWunderlich_etal_2021}, determining a team's strength is a difficult task because of the interaction between skills and randomness. Different approaches have been proposed to evaluate a team's strength, such as using the positions of teams in team rankings~\cite{BEvangelos_AGioldasis_GIoannis_AGeorgia_2018}, the total number of points scored in a competition~\cite{AHeuer_ORubner_2008}, ELO-ratings~\cite{LMHvattum_HArntzen_2010}, or betting odds~\cite{FWunderlich_etal_2021}. In this paper, weak teams are identified based on their positions in the weighted team ranking described in Section~\ref{sec:data_collection}.

%%%%%%%%%%%%%%%%%%%%%%%%%%%%%%%%%%%
\paragraph{Identifying weak teams.} 

Given the weighted team ranking in~\eqref{eq:weighted_team_ranking}, one can consider two strategies to identify weak teams. A first strategy consists of considering the top~$p\%$ teams in the weighted team ranking as strong and the bottom~$p\%$ teams as weak. However, in our case, this strategy proved unsuccessful because teams in the bottom~$p\%$ rarely, if ever, defeat teams in the top~$p\%$, regardless of the sport. When~$p=50$, for some sports, it occasionally happens that teams in the bottom half defeat teams in the top half, while in others, it never happens. Nevertheless, this is just noise occurring due to teams in mid-ranking positions, and so it is not a reliable indication of weak teams prevailing over strong ones. Therefore, we considered a second strategy, described below. 

The approach used in our paper consists of comparing the positions of two teams in the weighted team ranking 
% that aggregates past editions 
and considering as weak the team that is ranked significantly lower, if it exists. In other words, there must be a relatively high difference in positions between the teams in the weighted ranking to consider the lower-ranked team as weak. Using the notation introduced in Section~\ref{sec:data_collection}, for each edition~$e \in \mathcal{E}$,
we denote as~$c(i,{\mathcal{W}_{\leqslant e}}) \in [1,N_{\leqslant e}]$ the position of a team~$i \in \mathcal{P}_{\leqslant e}$ in the weighted team ranking~$\mathcal{W}_{\leqslant e}$. 
Given two teams~$i$ and~$j$ in~$\mathcal{P}_{\leqslant e}$, we refer to~$|c(i,{\mathcal{W}_{\leqslant e}}) - c(j,{\mathcal{W}_{\leqslant e}})|$ as the rank difference between teams~$i$ and~$j$ in the weighted team ranking~$\mathcal{W}_{\leqslant e}$. 
Given a match~$(i,j) \in \mathcal{M}_{e}$ between teams~$i$ and~$j$ in edition~$e$, we identify~$i$ as a weak team based on~$\mathcal{W}_{\leqslant e}$ if
\begin{equation}\label{eq:suff_decrease}
c(i,{\mathcal{W}_{\leqslant e}}) \; \le \; c(j,{\mathcal{W}_{\leqslant e}}) - \tau, 
\end{equation}
where~$\tau$ is a positive threshold depending on the sport, to be determined. Note that~\eqref{eq:suff_decrease} implies that the rank difference~$|c(i,{\mathcal{W}_{\leqslant e}}) - c(j,{\mathcal{W}_{\leqslant e}})|$ is greater than~$\tau$. 
% Now, recall~$\mathcal{E}_s = (e_1,e_2,\ldots,e_{|\mathcal{E}_s|})$, where~$e_1$ is the earliest available edition and~$e_{|\mathcal{E}_s|}$ is the most recent edition for sport~$s \in \mathcal{S}$. For any~$h \in \{2, \ldots, |\mathcal{E}_s|\}$, given a match~$(i,j) \in \mathcal{M}_{e_h}$ between teams~$i$ and~$j$ in edition~$e_h$, we identify~$i$ as a weak team based on~$\mathcal{W}_{\leqslant e_{h-1}}$ if
% \begin{equation}\label{eq:suff_decrease}
% c(i,{\mathcal{W}_{\leqslant e}} \; \le \; c(j,{\mathcal{W}_{\leqslant e}} - \tau, 
% \end{equation}
% where~$\tau$ is a positive threshold. Note that~\eqref{eq:suff_decrease} implies that the rank difference~$|wc_{\leqslant e_{h-1}}(i) - wc_{\leqslant e_{h-1}}(j)|$ is at least equal to~$\tau$. 
When~\eqref{eq:suff_decrease} is not satisfied and, therefore, two teams have a rank difference less than or equal to the threshold~$\tau$, we assume that such teams have similar strengths and we do not classify either of them as weak. 

%%%%%%%%%%%%%%%%%%%%%%%%%%%%%%%%%%%
\paragraph{Underdog achievement score.}

Recall~$\mathcal{E} = (e_1,e_2,\ldots,e_{|\mathcal{E}|})$, where~$e_1$ is the earliest available edition and~$e_{|\mathcal{E}|}$ is the most recent edition, and recall the definition of weak team based on~$e \in \mathcal{E}$ in~\eqref{eq:suff_decrease}. For any~$h \in \{2, \ldots, |\mathcal{E}|\}$, we define the underdog achievement score for edition~$e_h \in \mathcal{E}$ as follows
\begin{equation}\label{eq:UAS}
\begin{split}
\begin{aligned}
\UAS_{e_h} \; &= \; \frac{\text{Number of victories or draws in edition~$e_h$ by weak teams based on~$\mathcal{W}_{\leqslant e_{h-1}}$}}{\text{Number of matches in edition~$e_{h}$ with a weak team based on~$\mathcal{W}_{\leqslant e_{h-1}}$}}.\\
% \\[3ex]
% &= \frac{|\{(i,j) \in \mathcal{M}_{e_h} ~|~ \score_{ij}(i) \ge \score_{ij}(j) \text{, $i$ weak team based on~$\mathcal{W}_{\leqslant e_{h-1}}$}\}|}{|\mathcal{M}_{e_h}|}\\
% & \quad + \frac{|\{(i,j) \in \mathcal{M}_{e_h} ~|~ \score_{ij}(i) \ge \score_{ij}(j) \text{, $i$ weak team based on~$\mathcal{W}_{\leqslant e_{h-1}}$}\}|}{|\mathcal{M}_{e_h}|}\\
% & \quad + \frac{|\{(i,j) \in \mathcal{M}_{e_h} ~|~ \score_{ij}(i) = \score_{ij}(j) \text{, $i$ or $j$ weak team based on~$\mathcal{W}_{\leqslant e_{h-1}}$}\}|}{|\mathcal{M}_{e_h}|}.
\end{aligned}
\end{split}
\end{equation}
Note that in~\eqref{eq:UAS}, weak teams are identified using a weighted team ranking that incorporates all past editions except the current one to avoid biasing the results. One can interpret~\eqref{eq:UAS} as the probability that a historically weaker team wins against a historically stronger one in a certain edition of a competition. For each sport, the average underdog achievement score across all editions is given by
\begin{equation}\label{eq:avgUAS}
\UAS \; = \; \frac{1}{|\mathcal{E}|-1} \sum_{h=2}^{|\mathcal{E}|} \text{UAS}_{e_h}. 
\end{equation}

To investigate further definitions of underdog achievement, we will also consider the following alternative metric  
\begin{equation}\label{eq:avgUAS2}
\begin{split}
\begin{aligned}
\overline{\UAS} \; &= \; \frac{\cup_{h=2}^{|\mathcal{E}|} \{\text{No. of victories/draws in edition~$e_h$ by weak teams based on~$\mathcal{W}_{\leqslant e_{h-1}}$}\}}{\cup_{h=2}^{|\mathcal{E}|} \{\text{No. of matches in edition~$e_{h}$ with a weak team based on~$\mathcal{W}_{\leqslant e_{h-1}}$}\}},
\end{aligned}
\end{split}
\end{equation}
which aggregates the numerator and denominator of~$\UAS_{e_h}$ in~\eqref{eq:UAS} across all editions~$e_h$ in~$\mathcal{E}$, with~$h \in \{2, \ldots, |\mathcal{E}|\}$.

%%%%%%%%%%%%%%%%%%%%%%%%%%%%%%%%%%%
\paragraph{Numerical results.} 

In this subsection, we first perform a rank difference analysis to determine the value of the threshold~$\tau$, used to identify weak teams in~\eqref{eq:suff_decrease}, which affects the computation of~$\UAS_{e_h}$ and~$\UAS$ in~\eqref{eq:UAS} and~\eqref{eq:avgUAS}, respectively. Then, we quantify~$\UAS$ for each sport.
% The set~$\mathcal{E}_s$ consists of the editions in Table~\ref{tab:competitions_app} of Appendix~\ref{sec:app}.
For each edition~$e \in \mathcal{E}$,
we recall that~$c(i,{\mathcal{R}_e}) \in [1,N_e]$ denotes the position of a team~$i \in \mathcal{P}_e$ in the team ranking~$\mathcal{R}_e$. Given two teams~$i$ and~$j$ in~$\mathcal{P}_{e}$, we refer to~$|c(i,{\mathcal{R}_{e}}) - c(j,{\mathcal{R}_{e}})|$ as the rank difference between teams~$i$ and~$j$ in the team ranking~$\mathcal{R}_{e}$.

\begin{figure}%[H]
    \centering
          \includegraphics[scale=0.5]{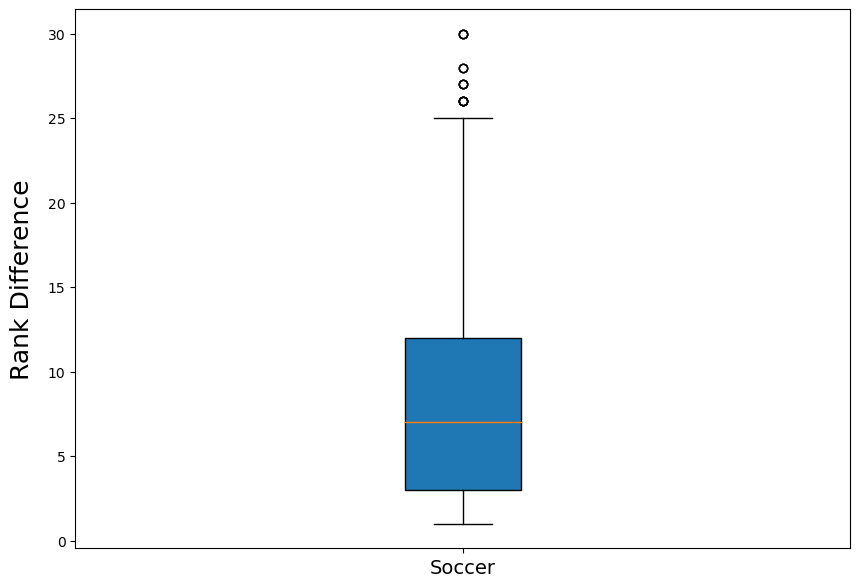}
        \caption{Box plot showing the distribution of rank differences between teams across all soccer matches.}\label{fig:boxplot_soccer}
\end{figure}

\begin{figure}%[H]
    \centering
          \includegraphics[scale=0.5]{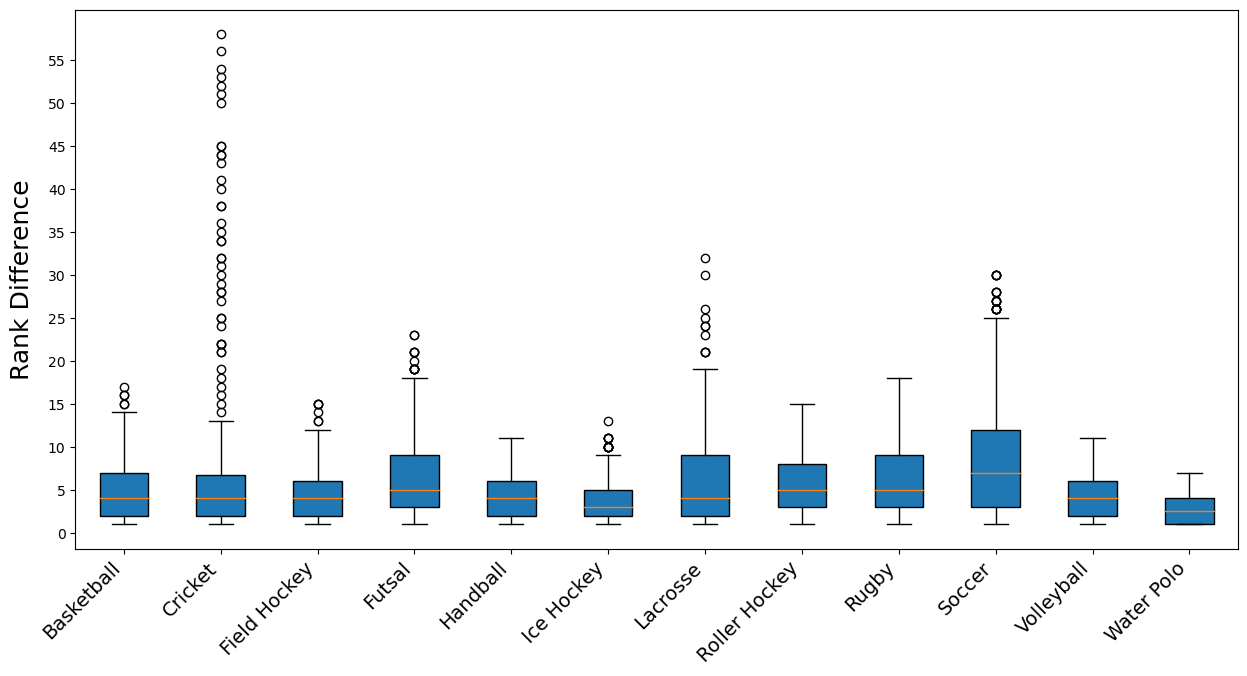}
        \caption{Box plot showing the distribution of rank differences between teams across all matches for each team ball sport included in our paper.}\label{fig:boxplot_sports}
\end{figure}

Figure~\ref{fig:boxplot_soccer} represents a box plot showing the distribution of the rank differences between teams~$i$ and~$j$ in the team ranking~$\mathcal{R}_e$ across all soccer matches for all editions, i.e.,~$|c(i,{\mathcal{R}_{e}}) - c(j,{\mathcal{R}_{e}})|$ for all~$(i,j) \in \mathcal{M}_e$ and~$e \in \mathcal{E}_{\text{soccer}}$ (where~$\mathcal{E}_{\text{soccer}}$ is the set of editions for soccer).
One can see that the rank differences range from~1 to~30, and approximately half of the soccer matches occurred with a rank difference less than or equal to~7, which corresponds to the median value of the distribution. Similarly, Figure~\ref{fig:boxplot_sports} represents box plots showing the distribution of the rank differences between teams~$i$ and~$j$ in the team ranking~$\mathcal{R}_e$ across all matches for all sports and editions, i.e.,~$|c(i,{\mathcal{R}_{e}}) - c(j,{\mathcal{R}_{e}})|$ for all~$(i,j) \in \mathcal{M}_e$, $e \in \mathcal{E}_s$, and~$s \in \mathcal{S}$ (where~$\mathcal{E}_s$ is the set of editions for sport~$s$). One can observe that the medians of the rank difference distributions across all sports range from~2.5 for water polo to~7 for soccer. In particular, for all sports except soccer, the median of the rank difference distribution is less than or equal to~5. 
% Similar observations can be repeated for the third quartiles of the rank difference distributions across all the sports, which range from~4 for water polo to~12 for soccer. 
For the rest of the paper, we set the threshold~$\tau$ in~\eqref{eq:suff_decrease} equal to the median of the corresponding rank difference distribution for each sport (see Table~\ref{tab:lambda_impact}).

\begin{table}
\centering
\begin{tabular}{|l|c|c|c|c|}
\cline{3-5}
\multicolumn{1}{c}{} & \multicolumn{1}{c}{} & \multicolumn{3}{|c|}{$\UAS$}\\
\hline
Sport & $\tau$ & $\lambda = 1$ & $\lambda = 0.5$ & $\lambda = 0$ \\
\hline
Basketball   & 4   & 0.25  & 0.19   & 0.16 \\
Cricket      & 4   & 0.15  & 0.11   & 0.08 \\
Field Hockey & 4   & 0.31  & 0.22   & 0.20 \\
Futsal       & 5   & 0.17  & 0.13   & 0.07 \\
Handball     & 4   & 0.21  & 0.17   & 0.11 \\
Ice Hockey   & 3   & 0.30  & 0.21   & 0.18 \\
Lacrosse     & 4   & 0.08  & 0.07   & 0.06 \\
Roller Hockey & 5  & 0.05  & 0.02   & 0.01 \\
Rugby        & 5   & 0.07  & 0.04   & 0.03 \\
Soccer       & 7   & 0.36  & 0.27   & 0.22 \\
Volleyball   & 4   & 0.22  & 0.11   & 0.07 \\
Water Polo   & 2.5 & 0.37  & 0.34   & 0.32 \\
\hline
\end{tabular}
\caption{Threshold~$\tau$ and impact of~$\lambda$ on the~$\UAS$ values for each team ball sport included in our paper.}
\label{tab:lambda_impact}
\end{table}

% Figure~\ref{fig:pchart} shows three proportion charts (or p-charts) representing the distribution of~UAS for all sports for three different values of the weight~$\lambda$ in~\eqref{eq:weighted_ranking}. A p-chart is a statistical control chart used to monitor the proportion of nonconforming items in a process. Control limits are included on the chart to help identify when the process is out of control, indicating that special causes may be present. In our case, the nonconfirming items are the sports with a significantly high or low~UAS. One can observe that regardless of the value of~$\lambda$, soccer is the sport with the highest~UAS and cricket, roller hockey, and rugby with the lowest~UAS. As the value of~$\lambda$ decreases, which indicates that past editions are given less weight in the weighted team ranking, the~UAS for lacrosse and water polo significantly increases, while the~UAS for futsal and volleyball significantly decreases. The UAS for all sports for three different values of~$\lambda$ are reported in Table~\ref{tab:lambda_impact}.

\begin{figure}
    \centering
          \includegraphics[scale=0.38]{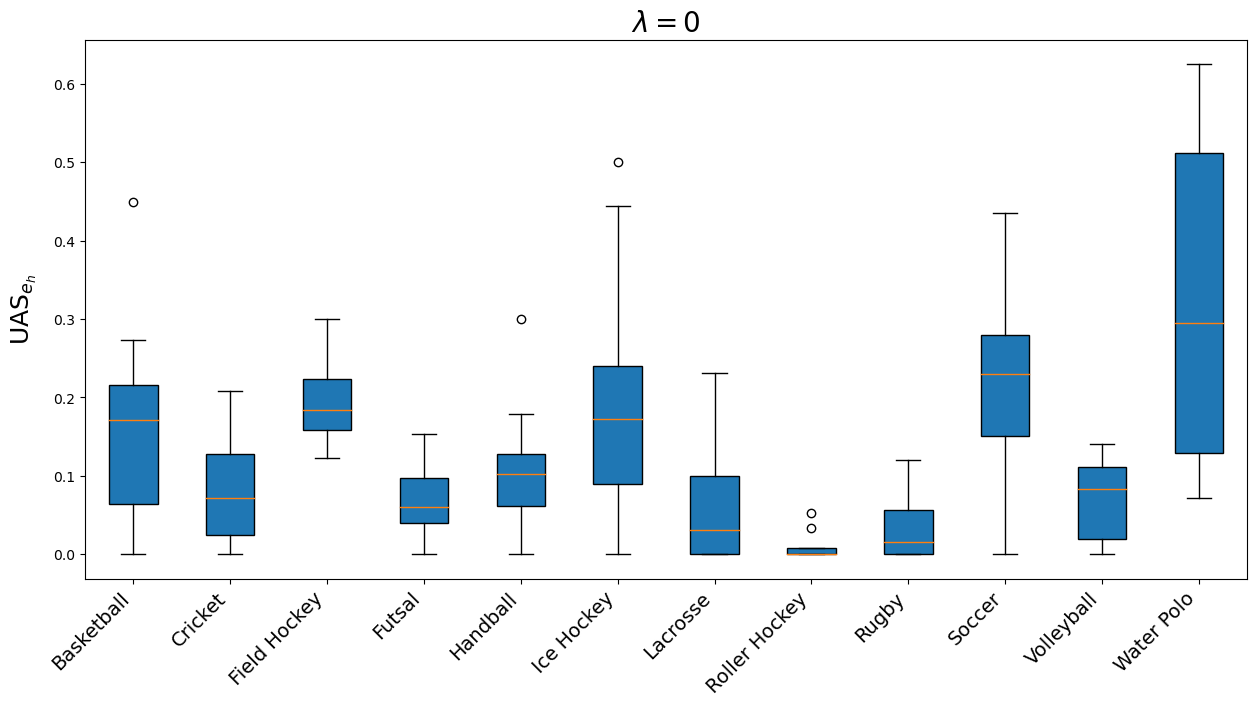}
          \includegraphics[scale=0.38]{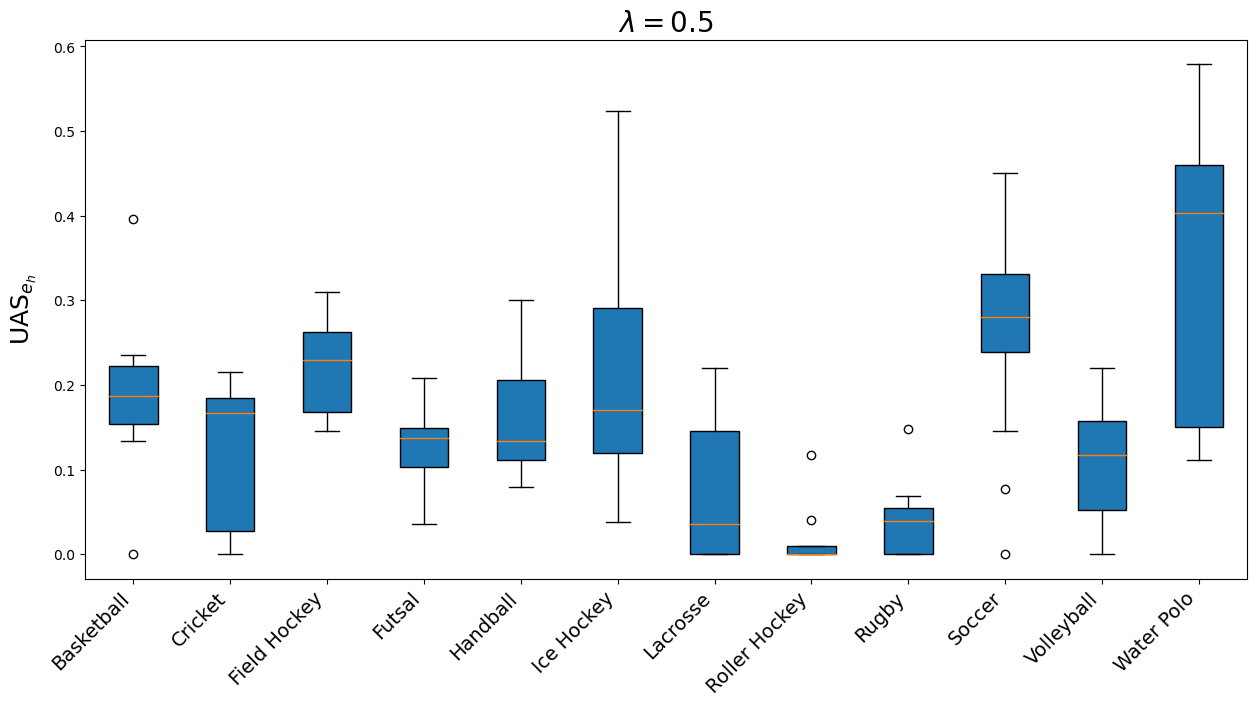}
          \includegraphics[scale=0.38]{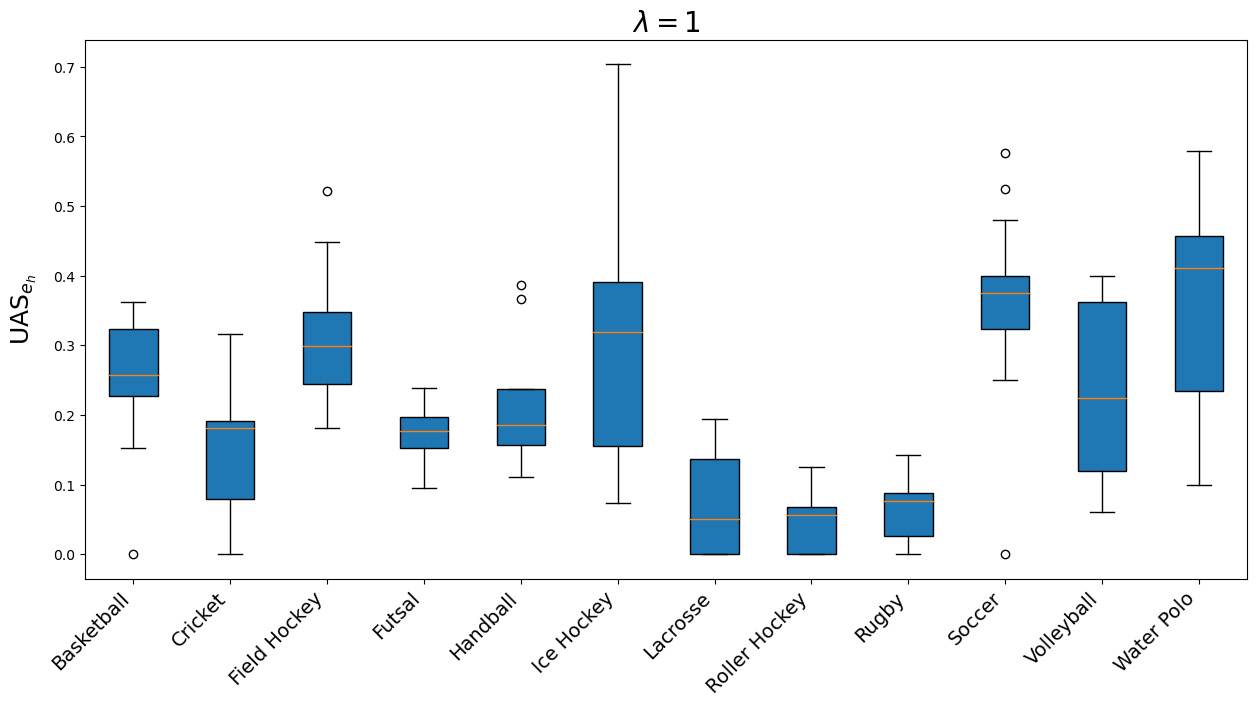}
        \caption{Box plots showing the~$\UAS_{e_h}$ distribution for each sport for different values of the weight~$\lambda$ in~\eqref{eq:weighted_ranking}.}\label{fig:pchart}
\end{figure}

\begin{figure}
    \centering
          \includegraphics[scale=0.47]{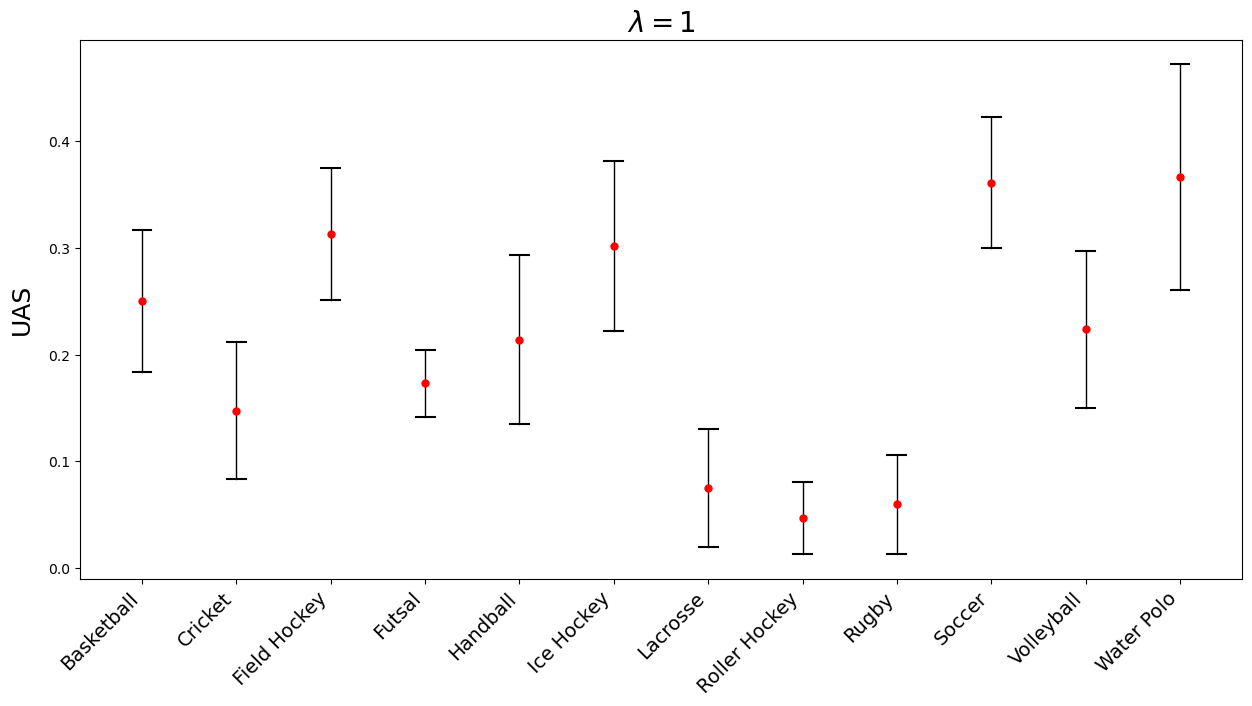}
        \caption{$95\%$-confidence intervals for the~$\UAS$ value for each sport for~$\lambda$ in~\eqref{eq:weighted_ranking} equal to~1.}\label{fig:conf_int}
\end{figure}

Figure~\ref{fig:pchart} contains three sets of box plots representing the distribution of~$\UAS_{e_h}$ values  across all editions~$e_h \in \mathcal{E}$ for each sport for three different values of the weight~$\lambda$ in~\eqref{eq:weighted_ranking}. One can observe that regardless of the value of~$\lambda$, field hockey, ice hockey, soccer, and water polo have consistently high values of~$\UAS_{e_h}$ compared to the other sports, while lacrosse, roller hockey, and rugby have consistently low values of~$\UAS_{e_h}$. Water polo is the sport for which the~$\UAS_{e_h}$ distribution has the highest median, followed by soccer. Conversely, lacrosse, roller hockey, and rugby are the sports for which the~$\UAS_{e_h}$ distribution has the lowest median.
% As the value of~$\lambda$ decreases, which indicates that past editions are given less weight in the weighted team ranking, the medians of the~$\UAS$ distributions for soccer and water polo approach each other. 
Similar results can also be observed for the average underdog achievement score in Figure~\ref{fig:conf_int}, which shows the~$\UAS$ value for each sport as a red point along with the corresponding confidence interval at a~95\% confidence level when~$\lambda$ in~\eqref{eq:weighted_ranking} is equal to~1. 
The~$\UAS$ values for each sport for three different values of~$\lambda$ are reported in Table~\ref{tab:lambda_impact}.

\begin{figure}
    \centering
          \includegraphics[scale=0.53]{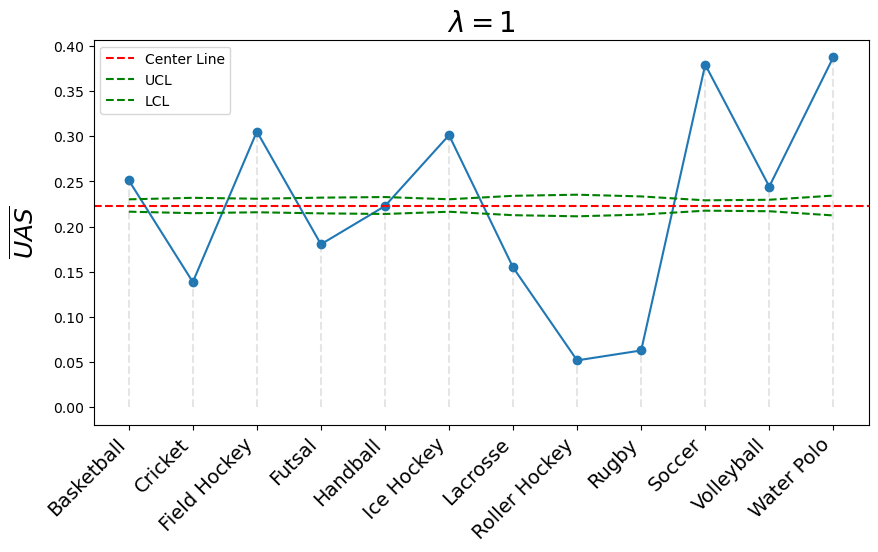}
        \caption{Laney p'-chart showing the~$\overline{\UAS}$ value for each sport for~$\lambda$ in~\eqref{eq:weighted_ranking} equal to~1.}\label{fig:laney_pchart}
\end{figure}

Figure~\ref{fig:laney_pchart} shows a Laney proportion chart (or Laney p'-chart) representing the~$\overline{\UAS}$ value for each sports when~$\lambda$ in~\eqref{eq:weighted_ranking} is equal to~1. Laney p'-charts are statistical control charts used to monitor the proportion of nonconforming items in a stochastic process~\cite{DBLaney_2002}. Control limits are included on the chart to help identify when the process is out of control, suggesting that special causes may be present.  
In our case, almost all sports fall outside the control limits, indicating that the~$\overline{\UAS}$ values across the sports do not follow the same distribution, which is expected since we are considering different sports. The reason why we decided to include this Laney p'-chart is to show that we obtain similar results in terms of underdog achievement regardless of the metric used, whether it is~\eqref{eq:UAS} in Figure~\ref{fig:pchart}, \eqref{eq:avgUAS} in Figure~\ref{fig:conf_int}, or~\eqref{eq:avgUAS2} in Figure~\ref{fig:laney_pchart}. In particular, from Figure~\ref{fig:laney_pchart}, one can observe that soccer and water polo have the highest~$\overline{\UAS}$ values, followed by~field hockey and ice hockey, while roller hockey and rugby have the lowest~$\overline{\UAS}$ values.

For simplicity, we will now perform statistical testing focusing on the case where~$\lambda$ in~\eqref{eq:weighted_ranking} is equal to~1. To statistically confirm the differences in the~$\UAS$ values across all the sports, we conducted a Kruskal-Wallis test~\cite{WHKruskal_WAWallis_1952}, which yielded a p-value of~$2.47 \cdot 10^{-10}$, indicating significant differences at a~5\% significance level. To identify the specific pairs of sports with statistically significant differences in their~$\UAS$ values, we resorted to Dunn's test with Bonferroni correction~\cite{ADinno_2015}, which performs multiple non-parametric pairwise comparisons among all the sports. The results of Dunn's test are included in Table~\ref{tab:stat_diff} only for pairs of sports with significant differences in their~$\UAS$ values at a~5\% significance level. Such results confirm the observations from Figures~\ref{fig:pchart}--\ref{fig:laney_pchart}, indicating that field hockey, ice hockey, soccer, and water polo exhibit higher underdog achievement compared to lacrosse, roller hockey, and rugby.

% \begin{table}[htbp]
% \centering
% \begin{tabular}{|l|l|l|l|}
% \cline{2-4}
% \multicolumn{1}{c}{} & \multicolumn{3}{|c|}{Dunn's test p-values}\\
% \cline{2-4}
% \multicolumn{1}{c|}{} & $\lambda=1$ & $\lambda = 0.5$ & $\lambda = 0$ \\
% \hline
% field hockey vs. lacrosse & 0.00884 & 0.08727 & 0.05025\\
% field hockey vs. roller hockey & 0.00244 & 0.00189 & 0.00066\\
% field hockey vs. rugby & 0.00841 & 0.01363 & 0.01108\\
% \hline
% ice hockey vs. lacrosse & 0.01292 & 0.61648 & 0.62142\\
% ice hockey vs. roller hockey & 0.00353 & 0.01589 & 0.01112\\
% ice hockey vs. rugby & 0.01294 & 0.10299 & 0.14606\\
% \hline
% soccer vs. cricket & 0.01239 & 0.13276 & 0.14249\\
% soccer vs. lacrosse & 0.00007 & 0.00352 & 0.01545\\
% soccer vs. roller hockey & 0.00002 & $3.60 \times 10^{-5}$ & $0.00012$ \\
% soccer vs. rugby & 0.00011 & 0.00045 & 0.00317\\
% \hline
% water polo vs. lacrosse & 0.00160 & 0.01819 & 0.03346\\
% water polo vs. roller hockey & 0.00044 & 0.00034 & 0.00047\\
% water polo vs. rugby & 0.00161 & 0.00271 & 0.00747\\
% \hline
% \end{tabular}
% \caption{Results of Dunn's test with Bonferroni correction for pairs of sports with significant differences in their~$\UAS_s$ values at a~5\% significance level.}\label{tab:stat_diff}
% \end{table}

\begin{table}[htbp]
    \centering
    \begin{tabular}{|l|l|}
        \cline{2-2}
        \multicolumn{1}{c|}{} & p-value \\
        \hline
        field hockey vs. lacrosse & 0.00884 \\
        field hockey vs. roller hockey & 0.00244 \\
        field hockey vs. rugby & 0.00841 \\
        \hline
        ice hockey vs. lacrosse & 0.01292 \\
        ice hockey vs. roller hockey & 0.00353 \\
        ice hockey vs. rugby & 0.01294 \\
        \hline
        soccer vs. cricket & 0.01239 \\
        soccer vs. lacrosse & $0.00007$ \\
        soccer vs. roller hockey & $0.00002$ \\
        soccer vs. rugby & 0.00011 \\
        \hline
        water polo vs. lacrosse & 0.00160 \\
        water polo vs. roller hockey & 0.00044 \\
        water polo vs. rugby & 0.00161 \\
        \hline
    \end{tabular}
    \caption{Results of Dunn's test with Bonferroni correction for pairs of sports with significant differences in their~$\UAS$ values at a~5\% significance level.}\label{tab:stat_diff}
\end{table}

% \begin{figure}
%     \centering
%           \includegraphics[scale=0.47]{Figs/PChart_1v1_wranking_decay1_linearRank}
%           \includegraphics[scale=0.47]{Figs/PChart_1v1_wranking_decay05_linearRank}
%           \includegraphics[scale=0.47]{Figs/PChart_1v1_wranking_decay0_linearRank}
%         \caption{P-Chart showing the UAS for each sport for different values of the weight~$\lambda$ in~\eqref{eq:weighted_ranking}.}\label{fig:pchart}
% \end{figure}

%%%%%%%%%%%%%%%%%%%%%%%%%%%%%%%%%%%%%%%%%%%%%%%
\section{Randomness model}\label{sec:randomness_model} 
%%%%%%%%%%%%%%%%%%%%%%%%%%%%%%%%%%%%%%%%%%%%%%%

In this section, we develop a model consisting of randomness factors that can affect match outcomes in team ball sports. In Section~\ref{sec:PCA_analysis}, such a model will be used to gain insights into the relationship between the randomness factors and underdog achievement. Unlike~\cite{FWunderlich_etal_2021} and~\cite{MLames_2018}, which propose variables of randomness affecting goal scoring in soccer as the match progresses, our model focuses on static factors, assuming scores as given. Therefore, we exclude factors that may influence player motivation, such as match location and current score, which are known to impact all sports and are not of interest to our analysis.

The factors that contribute to the inherent randomness observed in match outcomes are listed in Table~\ref{tab:factors} and categorized into three main groups: physical environment, player, and team. 
We believe that each of the factors in such a table should have a positive impact on randomness, meaning that a larger factor value corresponds to increased randomness.
To provide further clarity into the relationship between such factors and randomness, we will include explanations when describing the three groups of factors in more detail.
% Although the relationship between such factors and randomness should be quite intuitive, 
Table~\ref{tab:auxiliary_factors} includes companion factors used in the definition of some of the randomness factors in Table~\ref{tab:factors}.

For each sport, we quantified the average values of the randomness factors, resulting in a factors dataset containing~12 rows (one for each sport) and~14 columns (one for each factor). Such a factors dataset, presented in Table~\ref{tab:factors_dataset} of Appendix~\ref{sec:app_factors}, is provided in its normalized version. Table~\ref{tab:auxiliary_factors_dataset} serves as an auxiliary dataset used in the computation of values for Table~\ref{tab:factors_dataset}. 
For the sake of simplicity, we will use the term ``goals'' to collectively refer to various scoring targets, such as goals for soccer, baskets for basketball, and similar terms for the other team ball sports considered in our paper.
The values in the factors dataset in Table~\ref{tab:factors_dataset} were derived by first using the formulas defined in the next subsections of this section and then applying normalization to rescale the range of each column in~$[0,1]$. When applying normalization to each column, we used the formula
\[
a' \; = \; \frac{a - \min(a)}{\max(a) - \min(a)},
\]
where~$a'$ denotes the normalized value, $a$ denotes the original value, and~$\min(a)$ and~$\max(a)$ represent the minimum and maximum values that~$a$ takes on, respectively.

% Note that Table~\ref{tab:auxiliary_factors_dataset} includes~BW, PBP, PE, SF, and NTR, which are auxiliary quantities defined in Table~\ref{tab:auxiliary_factors}. 
% The values in the factors dataset in Table~\ref{tab:factors_dataset} were derived as follows. We first apply normalization to each column of the auxiliary dataset in Table~\ref{tab:auxiliary_factors_dataset} to obtain a normalized auxiliary dataset where the range of each column is~$[0,1]$. Then, to obtain the values of~BL, PBD, PI, SI, and~ATR in Table~\ref{tab:factors_dataset}, we use the formula~$1-a$, where~$a$ represents the values of the columns~BW, PBP, PE, SF, and NTR in the normalized auxiliary dataset. All the other columns in Table~\ref{tab:factors_dataset} are set equal to the corresponding columns in the normalized auxiliary dataset.

\begin{table}[h]
\centering
\begin{tabular}{|p{3.2cm}@{\hspace{0.5em}}p{12cm}|}
\hline
\multicolumn{2}{|c|}{\textbf{Physical Environment}}\\
\hline
BL: & Ball lightness\\
BV: & Ball velocity\\
FS/BS: & Field size/Ball size\\
GS/BS: & Goal size/Ball size\\
BG: & Ball geometry (Deviation from spherical form)\\
BB: & Ball bounciness\\
\hline
 \multicolumn{2}{|c|}{\textbf{Player}} \\
\hline
PP: & Player powerfulness\\
PBH: & Player ball handling\\
PBD: & Player ball dispossession\\
PI: & Player inexperience\\
\hline
 \multicolumn{2}{|c|}{\textbf{Team}} \\
\hline
NP/FS: & Number of players/Field size\\
GS/NPG: & Goal size/Number of players who can effectively defend the goal\\
SI: & Scoring infrequency\\
NRAM/NRPM: & Number of rules about movement/Number of rules that prevent movement\\
\hline
\end{tabular}
\caption{Factors that contribute to randomness in the match outcomes of the team ball sports included in our paper.}
\label{tab:factors}
\end{table}

\begin{table}[h]
\centering
\begin{tabular}{|p{3.2cm}@{\hspace{0.5em}}p{12cm}|}
\hline
% \multicolumn{2}{|c|}{\textbf{Physical Environment}}\\
% \hline
BW: & Ball weight\\
% \hline
%  \multicolumn{2}{|c|}{\textbf{Player}} \\
% \hline
PBP: & Player ball possession\\
PE: & Player experience\\
% \hline
%  \multicolumn{2}{|c|}{\textbf{Team}} \\
% \hline
SF: & Scoring frequency\\
\hline
\end{tabular}
\caption{Companion factors used in the definition of some of the randomness factors in Table~\ref{tab:factors}.}
\label{tab:auxiliary_factors}
\end{table}

% \begin{table}[h]
% \centering
% \begin{tabular}{|p{2cm}@{\hspace{0.5em}}p{12cm}|}
% \hline
% % \multicolumn{2}{|c|}{\textbf{Physical Environment}}\\
% % \hline
% BW: & Ball weight\\
% \hline
% %  \multicolumn{2}{|c|}{\textbf{Player}} \\
% % \hline
% PBP: & Player ball possession\\
% PE: & Player experience\\
% \hline
% %  \multicolumn{2}{|c|}{\textbf{Team}} \\
% % \hline
% SF: & Scoring frequency\\
% NTR: & Number of team rules that prevent movement\\
% \hline
% \end{tabular}
% \caption{Auxiliary quantities used to compute~BL, PBD, PI, SI, and~ATR in Table~\ref{tab:factors}.}
% \label{tab:auxiliary_factors}
% \end{table}

% \begin{table}[h]
% \centering
% \begin{tabular}{|l|}
% \hline
% \textbf{Physical Environment} \\
% \hline
% BW: Ball weight \\
% BV: Ball velocity \\
% BS/FS: Ball size/Field size \\
% BS/GS: Ball size/Goal size \\
% BG: Ball geometry (Deviation from sphere’s form) \\
% BB: Ball bounciness \\
% \hline
% \textbf{Player} \\
% \hline
% PP: Player powerfulness (Force on ball) \\
% PBH: Player ball handling (Proportion of body interacting with ball) \\
% PBP: Player ball possession (Actual play time divided by total number of players) \\
% PE: Player experience (Age) \\
% \hline
% \textbf{Team} \\
% \hline
% NP/FS: Number of players/Field size \\
% NPG/GS: Number of players who can effectively defend the goal/Goal size \\
% SF: Scoring frequency (Shots/Match time) \\
% NTR: Number of team rules that prevent movement \\
% \hline
% \end{tabular}
% \caption{Factors influencing sports randomness}
% \label{tab:factors}
% \end{table}

%%%%%%%%%%%%%%%%%%%%%%%%%%%%%%%%%%%
\paragraph{Physical environment factors.}

The physical environment category includes randomness factors related to properties of the sporting equipment and playing field. The formulas used to define such factors (including the units of measurement) are as follows: 
\vspace{0.2cm}

\renewcommand{\arraystretch}{1.2}
\begin{tabular}{ll}
BL & $\max(\text{BW}) - \text{BW}$, where BW is the ball weight~(gr) \\
BV & Average speed at which a player shoots the ball (km/h) \\
FS/BS & Surface of the field/Surface of the ball (m\textsuperscript{2}/m\textsuperscript{2}) \\
GS/BS & Surface of the goal/Surface of the ball (m\textsuperscript{2}/m\textsuperscript{2}) \\
BG & Categorical variable with three classes \\
& (0 for ice hockey, 2 for rugby, and~1 for all other sports) \\
BB & Categorical variable with eleven classes \\
& (from 0 for ice hockey to~11 for basketball).\\
\end{tabular}

\vspace{0.2cm}
Ball lightness~(BL), which is inversely related to ball weight~(BW), influences the force required for players to control a ball. We expect that lighter balls contribute more to randomness because they are generally more difficult to control due to their reduced mass, responding differently to player actions. Ball velocity~(BV) affects the timing of the gameplay, with faster balls expected to increase randomness. The ratio between field size and ball size~(FS/BS) is related to spatial dynamics. Higher values for such a ratio are associated with less ball control and more player movement, thus increasing randomness. 
The ratio between goal size and ball size~(GS/BS) influences the dynamics of a match in a similar way, as higher values for such a ratio imply a higher likelihood of scoring a goal and changing the match outcomes.
Since cricket lacks a goal, we estimate its ratio between goal size and ball size by averaging values obtained for other sports. Ball geometry~(BG), which refers to deviations from spherical form, introduces unpredictability in the paths a ball takes. We categorized such a factor into three classes (0 for ice hockey, 2 for rugby, and 1 for all other sports) to quantify its impact. Similarly, ball bounciness~(BB), which determines the extent of rebound upon impact, was treated as a categorical variable by assigning each sport to one of eleven categories, from~0 for ice hockey (no bounciness) to~11 for basketball (maximum bounciness). 

%%%%%%%%%%%%%%%%%%%%%%%%%%%%%%%%%%%
\paragraph{Player factors.}

The player category focuses on player attributes and skills that contribute to randomness. The formulas used to define such factors (including the units of measurement) are as follows: 
\vspace{0.2cm}

\renewcommand{\arraystretch}{1.2}
\begin{tabular}{ll}
PP & Body mass index = Weight/Height\textsuperscript{2} (kg/m\textsuperscript{2}) \\
PBH & Proportion of body interacting with the ball \\
PBD & $\max(\text{PBP}) - \text{PBP}$, where~PBP is the player ball possession, defined as follows:\\ 
& PBP = Actual play time/Total number of players on the field (minutes/players) \\
PI & $\max(\text{PE}) - \text{PE}$, where~PE is the player experience, defined as follows: \\
& PE = Average retirement age (years).\\
\end{tabular}

\vspace{0.2cm}
Player powerfulness~(PP) is related to the strength with which a player strikes a ball, influencing its trajectory and speed. Such a factor is measured in terms of the body mass index of a player, which is defined as the body mass divided by the square of the body height. Higher powerfulness is expected to increase randomness.
Player ball handling~(PBH) refers to the percentage of the body used to control a ball. In the case of cricket, lacrosse, field hockey, ice hockey, and roller hockey, such a percentage takes into account the sticks. Player ball dispossession~(PBD) refers to a player's inability to maintain possession of a ball and, therefore, is inversely related to player ball possession~(PBP), which we measure by the actual play time (i.e., match time, without including interruptions) divided by the total number of players on the field. Player ball dispossession influences scoring opportunities because the lower the possession, the lower the control on a ball, and the higher the contribution to randomness. 
Player inexperience~(PI) is inversely related to player experience~(PE), which we measure in terms of average retirement age. The average retirement age reflects the accumulation of skills and decision-making abilities over time, and thus results in performance consistency. Therefore, as the level of inexperience increases, so does the contribution to randomness.

%%%%%%%%%%%%%%%%%%%%%%%%%%%%%%%%%%%
\paragraph{Team factors.}

The team category includes randomness factors related to collective dynamics and match rules. The formulas used to define such factors (including the units of measurement) are as follows: 
\vspace{0.2cm} 

\renewcommand{\arraystretch}{1.2}
\begin{tabular}{ll}
NP/FS & Total number of players on the field/Surface of the field (players/m\textsuperscript{2}) \\
GS/NPG & Surface of the goal/Number of players who can defend the goal \\
& (m\textsuperscript{2}/players)\\
SI & $\max(\text{SF}) - \text{SF}$, where~SF is the scoring frequency, defined as follows:\\
& SF = No. of goals or points being scored per team/Actual play time\\
& (goals/min)\\
NRAM/NRPM & Ratio between the number of rules about movement and the number of \\
& rules that prevent movement, as defined in Table~\ref{tab:rule_description} of Appendix~\ref{sec:app_factors}. \\
\end{tabular}

\vspace{0.2cm}
The ratio between the number of players and the field size~(NP/FS) is a measure of the coverage of the field by players. A higher value of such a ratio is associated with a wider range of offensive and defensive strategies and, therefore, is expected to have a positive impact on randomness. The ratio between the goal size and the number of players who can effectively defend the goal~(GS/NPG) is a measure of the defensive weakness of a team. The fewer players defend the goal, the higher the variability in match outcomes. We estimate the value of such a ratio for cricket, which lacks a goal, by averaging the values obtained for the other sports. Scoring infrequency~(SI) refers to the low frequency with which goals are being scored during a match. Such a factor is inversely related to the scoring frequency~(SF), which we measure by the number of goals or points being scored per team divided by the actual play time. Sports with a low number of goals scored per match are more sensitive to randomness (in the sense of the final outcome of a match) and, therefore, the scoring infrequency can significantly impact the overall match outcome.
The presence of team rules that restrict movement imposes tactical constraints, limiting team dynamics and playstyle. Fewer movement constraints can result in greater unpredictability for match outcomes. Table~\ref{tab:rule_description} of Appendix~\ref{sec:app_factors} includes all the rules considered for the computation of the ratio between the number of rules about movement and the number of rules that prevent movement~(NRAM/NRPM).

% Basketball restricts movement through the 24-second shot clock and the prohibition of backcourt violations. Handball players face a three-second possession limit. Ice hockey enforces offsides to regulate player positioning. Lacrosse players are given a 20-second window to advance the ball beyond midfield. Rugby players must adhere to lateral or backward passing. Soccer enforces rules like offsides and penalties. Volleyball players are constrained by the three-hit rule, mandating that the ball cannot be hit more than three times before returning it to the opposing team. Water polo enforces a 30-second shot clock. Conversely, sports like cricket, futsal, roller hockey, and field hockey do not have any significant regulations regarding movement, allowing for a more fluid playstyle.

%%%%%%%%%%%%%%%%%%%%%%%%%%%%%%%%%%%%%%%%%%%%%%%
\section{Explaining the underdog achievement through our randomness model}\label{sec:PCA_analysis} 
%%%%%%%%%%%%%%%%%%%%%%%%%%%%%%%%%%%%%%%%%%%%%%%

In this section, we perform a principal component analysis~(PCA) and a correlation analysis to gain insights into the relationship between the~$\UAS$ computed for each sport in Section~\ref{sec:underdog_achievement} (see Table~\ref{tab:lambda_impact}) and the randomness factors introduced in Table~\ref{tab:factors} of Section~\ref{sec:randomness_model} and quantified in the factors dataset in Table~\ref{tab:factors_dataset} of Appendix~\ref{sec:app_factors}. PCA is a statistical technique used to simplify a dataset by computing a linear combination of its column values (one can think of a column as a vector) while preserving the variability in the original dataset~\cite{ITJolliffe_JCadima_2016}. PCA plots can be used to visually represent the results of the analysis through scatterplots, and these are the types of plots that we use to determine the relative importance of the randomness factors for each sport. A correlation analysis will then be performed to study the linear correlation between each pair of factors, including the underdog achievement.

%%%%%%%%%%%%%%%%%%%%%%%%%%%%%%%%%%%
\paragraph{Principal component analysis.}

\begin{figure}
    \centering
          \includegraphics[scale=0.55]{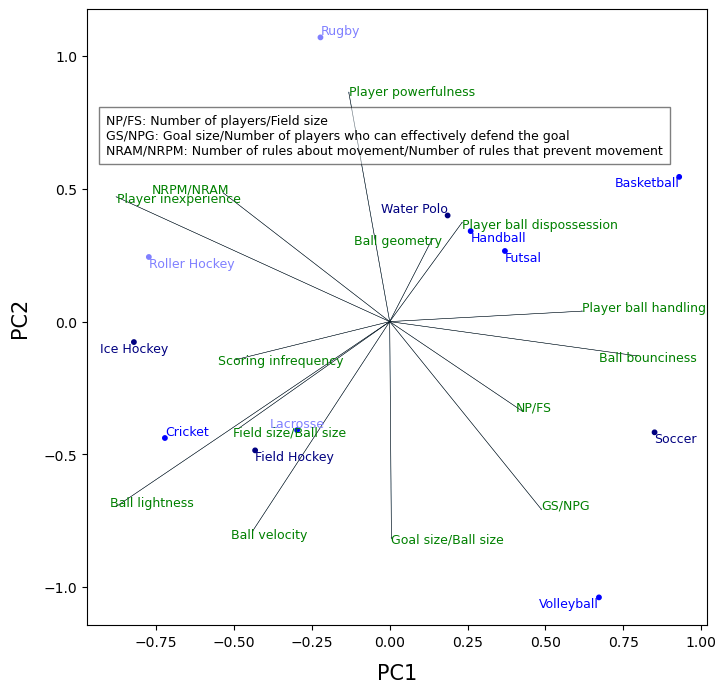}
          \includegraphics[scale=0.55]{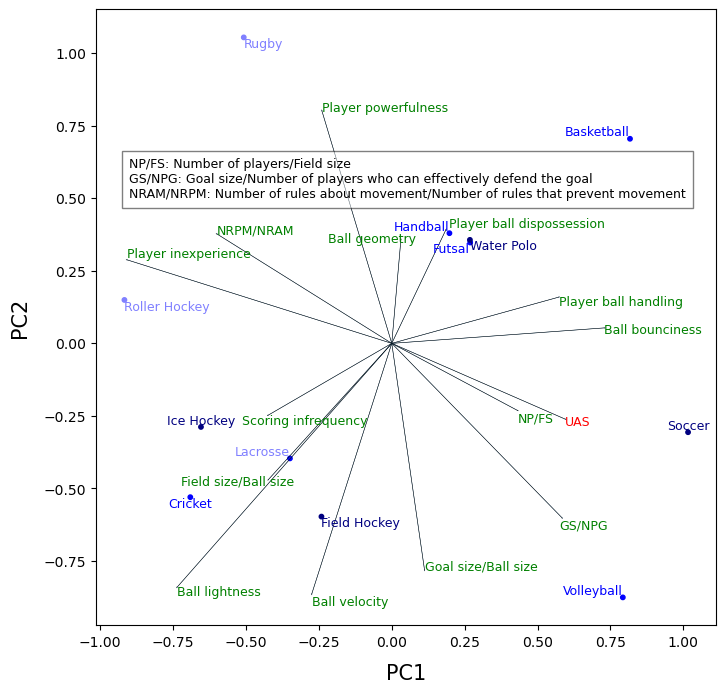}
        \caption{PCA plots, without~$\UAS$ (upper plot) and with~$\UAS$ (lower plot).}\label{fig:PCA_plot}
\end{figure}

\begin{figure}
    \centering
          \includegraphics[scale=0.40]{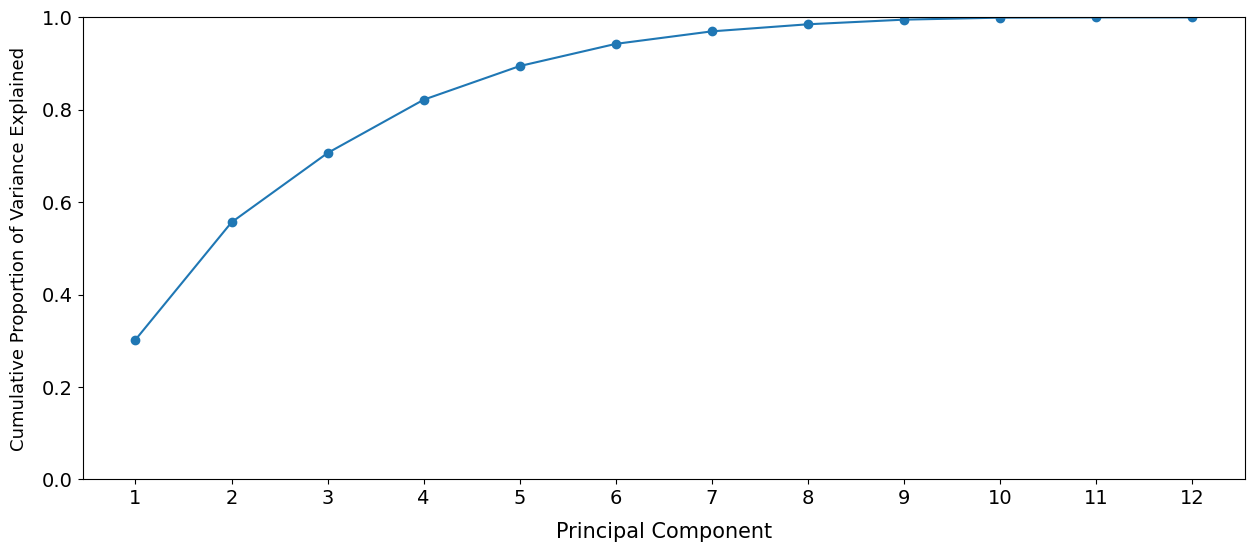}
        \caption{Scree plot showing the cumulative proportion of variance explained by each additional principal component when~$\UAS$ is not included among the factors. The scree plot when~$\UAS$ is included is almost identical and is omitted.}\label{fig:scree_plot}
\end{figure}

By applying~PCA, one can transform the factors dataset in Table~\ref{tab:factors_dataset}, consisting of~14 columns, into a reduced dataset with as many columns as the number of principal components selected, where each principal component is obtained as a linear combination of the columns in the original factors dataset. Figure~\ref{fig:PCA_plot} represents the~PCA plots for the first two principal components resulting from the application of~PCA to the original factors dataset (upper plot) and the factors dataset with an additional column consisting of the~$\UAS$ values obtained from Table~\ref{tab:lambda_impact} in Section~\ref{sec:underdog_achievement} when~$\lambda= 1$ (lower plot). 
The~PCA plots in Figure~\ref{fig:PCA_plot} provide a graphical representation of the relationships between sports and factors captured by the first two principal components. 
In both~PCA plots, the data points, represented as blue dots, are associated with the sports. We use different shades of blue to denote whether their~$\UAS$ is high (dark blue), medium (medium blue), or low (light blue), based on the results described in Section~\ref{sec:underdog_achievement}. The coordinates of each data point are determined by the so-called scores, which are new variables associated with the columns in the reduced dataset. 
% Each principal component corresponds to a direction in the multidimensional space defined by the original~14 factors, and the scores represent the projection of the data points onto these directions.
In general, the first two principal components are the ones that capture the maximum variance in a dataset. In our case, they are able to explain~56\% of the variability in the original factors dataset, as shown in the scree plot in Figure~\ref{fig:scree_plot}. 
In addition to the data points associated with the sports, the~PCA plots include vectors, referred to as loadings and depicted as line segments, which represent the contribution of each original factor to the variability in the factors dataset explained by the principal components. 
% Each loading vector indicates the direction and magnitude of the corresponding factor's influence on the principal components. 
Factors associated with loading vectors with similar directions and magnitudes have similar importance in explaining the variability. In the~PCA plots in Figure~\ref{fig:PCA_plot}, the magnitude of each loading vector has been doubled for improved clarity.

 % By looking at the directions of the loading vectors in the~PCA plots in Figure~\ref{fig:PCA_plot}, one can observe the relative importance of the original factors in explaining the variability observed in the entire factors dataset. 
 
 By examining the positions of the data points associated with the sports and the directions and magnitudes of the loading vectors in each of the~PCA plots in Figure~\ref{fig:PCA_plot}, one can observe the relative importance of the randomness factors for each sport.
 
 \textbf{Observation 1.} 
In soccer, PBH, BB, and~GS/NPG exhibit high values because soccer players are allowed to use various body parts to interact with the ball, the ball tends to be highly bouncy, and there is only one player who defends the goal. We note that although~NP/FS is close to soccer (in the sense of the plot), soccer actually has one of the smallest values for such a ratio. The close proximity of~NP/FS to soccer is perhaps due to the close proximity of soccer to volleyball in the plot, which has a high value for such a ratio. 
 % and because~PCA only accounts for~50\% of the variability in the data, as mentioned earlier. 

 \textbf{Observation 2.}
 For the hockey sports (i.e., field hockey, ice hockey, and roller hockey), one can observe that the predominant randomness factors are~PI, SI, FS/BS, BL, and~BV. Indeed, in these sports, players retire at a relatively young age, scoring frequency is lower compared to other sports like basketball, ball sizes are small (resulting in a high~FS/BS value), ball weight is light, and ball velocity is high. 
 
 \textbf{Observation 3.}
 For water polo, the main randomness factors are~PBD and~PP. The high~PBD value is due to the low actual play time compared to other sports. Similar conclusions can be extended to handball, futsal, and basketball, which are all close to water polo (in the sense of the plot). 

 \textbf{Observation 4.}
 For rugby, the most significant randomness factors are~PP, BG, PI, and~PBD, each attaining their maximum values. BG has a high value due to the unique shape of rugby balls, which contributes to increased randomness in match outcomes. PI has a high value because rugby players typically retire at a young age. The high~PBD value is due to a combination of very low actual play time and a very high number of players. NRAM/NRPM also takes on a high value due to the fewer movement restrictions in rugby compared to other sports.
 
 % Ball weight, number of team rules that prevent movement, ball size/goal size, scoring frequency, number of players who can effectively defend the goal/goal size, and ball velocity have the highest influence on the first principal component (note that most of such factors belong to the physical environment and team categories of Table~\ref{tab:factors}). Instead, player powerfulness, player ball handling, ball geometry, player ball possession, ball size/field size, number of players/field size, and player experience have the highest influence on the second principal component (note that most of such factors belong to the player category of Table~\ref{tab:factors}). All the other factors contribute approximately equally to the two principal components. These results suggest that factors within the physical environment and team categories contribute the most to the sports variability in the factors dataset, while factors within the player category are responsible for the second-most variability. 
 % Furthermore, the lower~PCA plot indicates that the contribution to sports variability from~UAS is similar to that from player experience, number of players/field size, ball size/field size, and ball bounciness. 
 
 %%%%%%%%%%%%%%%%%%%%%%%%%%%%%%%%%%%
\paragraph{Correlation analysis.}
 
\begin{figure}
    \centering
          \includegraphics[scale=0.55]{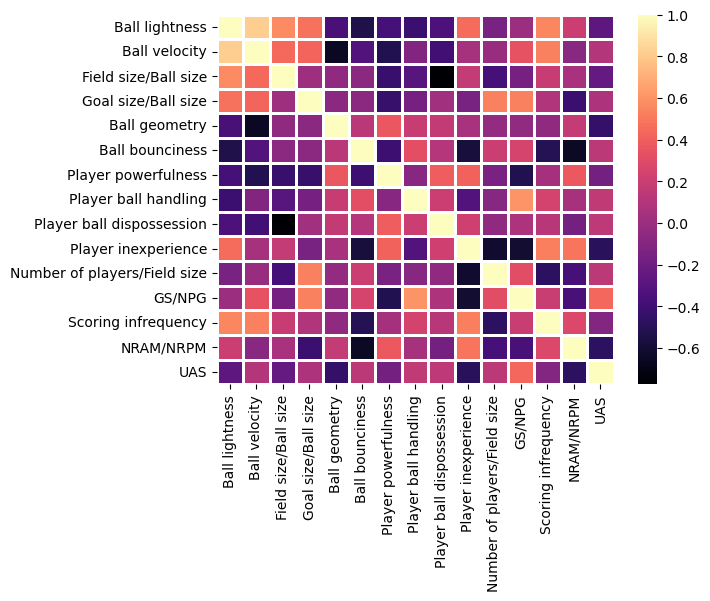}
        \caption{Heatmap illustrating the correlation between each pair of factors, including $\UAS$.}\label{fig:heatmap}
\end{figure}

 To gain insights into the relationship between underdog achievement and randomness factors, we report Figure~\ref{fig:heatmap}, which shows a heatmap illustrating the Pearson correlation coefficient between each pair of factors, including~$\UAS$. In general, the Pearson correlation coefficient between two variables is given by the covariance of two variables divided by the product of their standard deviations, and it is a measure of their linear correlation, ranging from~-1 (negative correlation) to~1 (positive correlation). The heatmap in Figure~\ref{fig:heatmap} indicates that~$\UAS$ exhibits the strongest positive correlation with~GS/NPG and the strongest negative correlation with~NRAM/NRPM, PI, and~BG. Additionally, $\UAS$ exhibits a relatively weaker negative correlation with~SI, PP, FS/BS, and~BL. The positive correlation between~$\UAS$ and~GS/NPG is expected, as~GS/NPG takes on a high value in soccer, which is among the sports with the highest~$\UAS$. The negative correlations observed with~$\UAS$ may appear surprising because they suggest that higher values of the factors decrease the underdog achievement, while we would expect that higher values of the factors increase randomness in match outcomes. However, this behavior is expected, as not all randomness factors have the same influence on underdog achievement. One can interpret the factors exhibiting a negative correlation with~$\UAS$ as having a weaker effect on randomness than the other factors. For example, rugby has a high value for~PI, which is negatively correlated with~$\UAS$. One can interpret this as the fact that the contribution of~PI to randomness in match outcomes is lower compared to other sports, which results in a low underdog achievement for rugby. By adopting such an interpretation, we conclude that the factors with the highest impact on randomness are those that exhibit a positive correlation with underdog achievement, i.e.,~GS/NPG, NP/FS, PBD, PBH, BB and, to a lesser extent, GS/BS, and~BV. 

% The PCA plots in Figure~\ref{fig:PCA_plot} also enable the visual identification of clusters of sports. Based on the positions of the data points associated with the sports, three distinct clusters can be observed. The first cluster comprises futsal, handball, soccer, volleyball, and water polo, which exhibit medium to high~UAS values when~$\lambda$ in~\eqref{eq:weighted_ranking} is equal to~1. The second cluster includes cricket, field hockey, ice hockey, lacrosse, and roller hockey, characterized by medium to low~UAS values. When~UAS is included among the factors, rugby is also included in the second cluster. Finally, the third cluster consists solely of basketball, which has a medium~UAS. These findings were further confirmed by performing a~$k$-means cluster analysis with~$k=3$ on both the original dataset and the reduced dataset consisting of the first two principal components. 

%%%%%%%%%%%%%%%%%%%%%%%%%%%%%%%%%%%%%%%%%%%%%%%
\section{Concluding remarks and future work}\label{sec:conclusions} 
%%%%%%%%%%%%%%%%%%%%%%%%%%%%%%%%%%%%%%%%%%%%%%%

In this paper, we studied the relationship between underdog achievement and randomness factors that affect match outcomes in team ball sports. To achieve our goal, we collected a vast amount of data containing information related to match scores, and we computed corresponding team rankings for each edition of the competitions selected for each sport. Then, we developed an underdog achievement score to determine the sports with the highest and lowest occurrences of weaker teams defeating stronger ones. Our findings indicate that water polo, soccer, field hockey, and ice hockey are among the sports with the highest underdog achievement, while lacrosse, roller hockey, and rugby are the ones with the lowest underdog achievement. Subsequently, we designed a randomness model consisting of~14 factors that contribute to unexpected match outcomes within each sport, providing quantitative values for each of the factors.
Finally, we performed~PCA and correlation analysis demonstrating that our randomness model can explain the underdog achievement. The randomness factors with the highest impact on underdog achievement are the ratio between the goal size and the number of players who can effectively defend the goal, the ratio between the number of players and the field size, player ball dispossession, player ball handling, and ball bounciness.   

For future research, we plan to replicate the analysis using competitions organized within professional sports leagues as well as collegiate sports competitions. Additionally, we aim to investigate the applicability of the methodology to team non-ball sports.

%%%%%%%%%%%%%%%%%%%%%%%%%%%%%%%%%%%%%%

\section*{Acknowledgments}
This work is partially supported by the U.S. Air Force Office of Scientific Research (AFOSR) award FA9550-23-1-0217.

%\bibliography{ref-sports}

\appendix

\section{International competitions for the considered team ball sports}\label{sec:app}
Table~\ref{tab:competitions_app} contains the major international competitions and the corresponding edition years selected for the team
ball sports included in our paper.

\begin{table}[b]
\centering
\begin{tabular}{|l|l|l|}
\hline
\textbf{Sport} & \textbf{Competition} & \textbf{Edition Years} \\
\hline
Basketball & Summer Olympic Games & 1964, 1968, 1972, 1976, 1980, 1984, \\
& & 2000, 2004, 2008, 2012, 2016, 2020\\\hline
Cricket & ICC Men's Cricket World Cup & 1975, 1979, 1983, 1987, 1992, 1996, \\
& & 1999, 2003, 2007, 2011, 2015, 2019\\\hline
Field Hockey & Men's FIH Hockey World Cup & 1971, 1973, 1975, 1978, 1982, 1986, \\
& & 1994, 1998, 2002, 2006, 2010, 2018,\\
& & 2023\\\hline
Futsal & FIFA Futsal World Cup & 1989, 1992, 1996, 2000, 2004, 2008, \\
& & 2012, 2016, 2020 \\\hline
Handball & Summer Olympic Games & 1976, 1988, 1992, 1996, 2000, 2008, \\
& & 2012, 2016, 2020 \\\hline
Ice Hockey & Winter Olympic Games & 1924, 1928, 1932, 1948, 1952, 1956, \\
& & 1960, 1964, 1972, 1976, 1980, 1984, \\
& & 1988, 2002, 2006, 2010, 2014, 2018,\\
& & 2022\\\hline
Lacrosse & World Lacrosse Men's World Cup & 1974, 1978, 1982, 1986, 1990, 1994, \\
& & 1998, 2002, 2006, 2010, 2014 \\\hline
Roller Hockey & World Skate Roller Hockey World Cup & 1999, 2001, 2003, 2005, 2007, 2009, \\
& & 2011, 2013, 2015 \\\hline
Rugby & Rugby World Cup &  1987, 1991, 1995, 1999, 2003, 2007,\\
& & 2011, 2015, 2019 \\\hline
Soccer & FIFA World Cup & 1930, 1934, 1950, 1954, 1958, 1962, \\
& & 1966, 1970, 1974, 1978, 1982, 1986,\\
& & 1990, 1994, 1998, 2002, 2006, 2010, \\
& & 2014 \\\hline
Volleyball & FIVB Volleyball Men's World Cup & 1965, 1969, 1977, 1981, 1985, 1989, \\
& & 1991, 1995, 1999, 2003, 2007, 2011, \\
& & 2015, 2019\\\hline
Water Polo & FINA Men's Water Polo World Cup & 1979, 1981, 1983, 1985, 1993, 1995, \\
& & 1999, 2002, 2010, 2014, 2018\\
\hline
\end{tabular}
\caption{Major international competitions and corresponding edition years selected for the team ball sports included in our paper.}
\label{tab:competitions_app}
\end{table}

\section{Factors dataset}\label{sec:app_factors}
Table~\ref{tab:factors_dataset} is the factors dataset described in Section~\ref{sec:randomness_model}, provided in its normalized version. Table~\ref{tab:auxiliary_factors_dataset} serves as an auxiliary dataset used in the computation of values for Table~\ref{tab:factors_dataset}. Table~\ref{tab:rule_description} summarizes the rules about movement and those preventing movement used in the computation of~NRAM/NRPM in Table~\ref{tab:auxiliary_factors_dataset}.

% \clearpage

\begin{sidewaystable}[h]
\footnotesize	
\centering
\begin{tabular}{|l|c|c|c|c|c|c|c|c|c|c|c|c|c|c|c|}
\hline
\textbf{Sports} & \textbf{BL} & \textbf{BV} & \textbf{FS/BS} & \textbf{GS/BS} & \textbf{BG} & \textbf{BB} & \textbf{PP} & \textbf{PBH} & \textbf{PBD} & \textbf{PI} & \textbf{NP/FS} & \textbf{GS/NPG} & \textbf{SI} & \textbf{NRAM/NRPM} \\
\hline
Basketball & 0 & 0 & 0.0013 & 0 & 0.5 & 1 & 0.270 & 0.057 & 0.802 & 0.25 & 0.30 & 0 & 0 & 0.17 \\
Cricket & 0.97 & 0.80 & 1 & 0.066 & 0.5 & 0.4 & 0 & 0 & 0.52 & 0.55 & 0 & 0.073 & 0.83 & 0.53 \\
Field Hockey & 0.97 & 0.61 & 0.29 & 0.79 & 0.5 & 0.3 & 0.012 & 0 & 0.92 & 0.89 & 0.041 & 0.42 & 0.98 & 0.17 \\
Futsal & 0.40 & 0.41 & 0.0052 & 0.062 & 0.5 & 0.7 & 0.193 & 1 & 0.85 & 0.69 & 0.15 & 0.32 & 0.95 & 0.53 \\
Handball & 0.33 & 0.41 & 0.0061 & 0.074 & 0.5 & 0.7 & 0.431 & 0.057 & 0.94 & 0.60 & 0.22 & 0.024 & 0.82 & 0 \\
Ice Hockey & 0.96 & 1 & 0.11 & 0.23 & 0 & 0 & 0.486 & 0 & 0.79 & 0.91 & 0.079 & 0.10 & 0.99 & 0.42 \\
Lacrosse & 1 & 0.74 & 0.46 & 0.43 & 0.5 & 0.9 & 0.012 & 0 & 0.94 & 0.92 & 0.031 & 0.17 & 0.92 & 0 \\
Roller Hockey & 0.98 & 0.59 & 0.056 & 0.164 & 0.5 & 0.1 & 0.259 & 0 & 0.82 & 0.92 & 0.12 & 0.077 & 0.93 & 1 \\
Rugby & 0.37 & 0.089 & 0.042 & 0.126 & 1 & 0.2 & 1 & 0.32 & 0 & 1 & 0.029 & 0.039 & 0.93 & 0.67 \\
Soccer & 0.38 & 0.67 & 0.070 & 0.188 & 0.5 & 0.8 & 0.009 & 1 & 0.91 & 0.083 & 0.0096 & 1 & 1 & 0.17 \\
Volleyball & 0.73 & 0.74 & 0 & 1 & 0.5 & 0.6 & 0.020 & 0.19 & 0.65 & 0 & 1 & 0.75 & 0.59 & 0 \\
Water Polo & 0.39 & 0.35 & 0.0013 & 0.0095 & 0.5 & 0.5 & 0.463 & 0.057 & 0.94 & 0.5 & 0.30 & 0.13 & 0.77 & 0.3 \\
\hline
\end{tabular}
\caption{Normalized factors dataset that quantifies the values of the randomness factors described in Section~\ref{sec:randomness_model} for each of the team ball sports included in our paper.}
\label{tab:factors_dataset}
\end{sidewaystable}

%%%%%%%%%%%

\clearpage

\begin{sidewaystable}[h]
\scriptsize
\centering

\begin{tabular}{|l|c|c|c|c|c|c|c|c|c|c|c|c|c|c|c|}
\hline
\textbf{Sports} & \textbf{BW} & \textbf{BV} & \textbf{FS/BS} & \textbf{GS/BS} & \textbf{BG} & \textbf{BB} & \textbf{PP} & \textbf{PBH} & \textbf{PBP} & \textbf{PE} & \textbf{NP/FS} & \textbf{GS/NPG} & \textbf{SF} & \textbf{NRAM/NRPM} \\
& {\scriptsize (g)} & {\scriptsize (km/h)} & {\scriptsize(m\textsuperscript{2}/m\textsuperscript{2})} & {\scriptsize(m\textsuperscript{2}/m\textsuperscript{2})} &  &  & {\scriptsize(kg/$\text{m}^2$)} &  & {\scriptsize(min/pl.)} & {\scriptsize(years)} & {\scriptsize (pl./$\text{m}^2$)} & {\scriptsize ($\text{m}^2$/pl.)} & {\scriptsize (goals/min)} & \\
\hline
Basketball & 602 & 29 & $2.4\cdot10^3$ & 119.1 & 1 & 10 & 24.78 & 0.092 & 4.8 & 33 & 0.023 & 0.45 & 2.4 & 1.29 \\
Cricket & 159.5 & 128 & $1.0\cdot10^6$ & NA & 1 & 4 & 23.15 & 0.054 & 19.09 & 29.39 & 0.0014 & NA & NA & 1.6 \\
Field Hockey & 160 & 104 & $3.\cdot10^5$ & 460.6 & 1 & 3 & 23.22 & 0.054 & 2.73 & 25.37 & 0.0044 & 7.83 & 0.073 & 1.29 \\
Futsal & 420 & 80 & $6.3\cdot10^3$ & 47.6 & 1 & 7 & 24.31 & 0.72 & 4 & 27.76 & 0.012 & 6 & 0.15 & 1.6 \\
Handball & 450 & 79.2 & $7.2\cdot10^3$ & 54.1 & 1 & 7 & 25.74 & 0.092 & 2.29 & 28.8 & 0.018 & 0.86 & 0.46 & 1.14 \\
Ice Hockey & 163 & 152.5 & $1.1\cdot10^5$ & 144 & 0 & 0 & 26.08 & 0.054 & 5 & 25.1 & 0.0071 & 2.16 & 0.12 & 1.5 \\
Lacrosse & 145 & 121 & $4.6\cdot10^5$ & 257.6 & 1 & 9 & 23.22 & 0.054 & 2.4 & 25 & 0.0036 & 3.35 & 0.21 & 1.14 \\
Roller Hockey & 155 & 102 & $5.7\cdot10^4$ & 105 & 1 & 1 & 24.71 & 0.054 & 4.5 & 25 & 0.01 & 1.79 & 0.2 & 2 \\
Rugby & 435 & 40 & $4.3\cdot10^4$ & 84 & 2 & 2 & 29.17 & 0.27 & 1.27 & 24 & 0.0035 & 1.12 & 0.2 & 1.71 \\
Soccer & 430 & 112 & $7.1\cdot10^4$ & 119.1 & 1 & 8 & 23.20 & 0.72 & 2.91 & 35 & 0.0021 & 17.86 & 0.03 & 1.29 \\
Volleyball & 270 & 121 & $1.2\cdot10^3$ & 578.6 & 1 & 6 & 23.27 & 0.18 & 7.5 & 36 & 0.074 & 13.5 & 1 & 1.14 \\
Water Polo & 425 & 72 & $2.5\cdot10^3$ & 17.8 & 1 & 5 & 25.93 & 0.092 & 2.29 & 30 & 0.023 & 2.7 & 0.56 & 1.4 \\
\hline
\end{tabular}
\caption{Auxiliary dataset used in the computation of values for Table~\ref{tab:factors_dataset}. For quantities that are not adimensional, we include the corresponding unit of measurement below the name of the column.}
\label{tab:auxiliary_factors_dataset}

\end{sidewaystable}

\clearpage

{\scriptsize
\begin{longtable}{|l|p{12cm}|}
    \hline
    \textbf{Sport} & \textbf{Rules about movement (RAM) and rules preventing movement (RPM)} \\
    \hline
    \endfirsthead
    
    % \multicolumn{2}{c}%
    % {{\tablename\ \thetable{} -- continued from previous page}} \\
    \hline
    % \textbf{Sport} & \textbf{Description of rules about movement} \\
    % \hline
    \endhead

    \hline
    %\multicolumn{2}{|r|}{{Continued on next page}} \\ \hline
    \endfoot
    
    \hline
    \caption{Table that summarizes the rules about movement~(RAM) and those preventing movement~(RPM) used in the computation of~NRAM/NRPM in Table~\ref{tab:auxiliary_factors_dataset}.}\label{tab:rule_description}
    \endlastfoot
    
    Basketball 
        &  1. Traveling: Restricts players from taking steps without dribbling the ball.\\
        (RAM: 1--9) & 2. Dribbling: Governs the legal handling of the ball while moving.\\
        (RPM: 1--7) & 3. Charging and blocking: Regulates contact between offensive and defensive players.\\
        & 4. Screening and blocking: Controls the legality of screens and blocking movements.\\
        & 5. 3-Second violation: Limits offensive players' time spent in the key area. \\
        & 6. 5-Second violation: Restricts the time allowed to inbound or shoot free throws.\\
        & 7. Out-of-bounds: Determines possession and restarts play when the ball goes out of bounds.\\
        & 8. Illegal contact: Penalizes players for illegal physical contact with opponents.\\
        & 9. Offensive foul for pushing off: Prohibits offensive players from pushing off. \\
    \hline
    Cricket  
        & 1. Running between the wickets: Regulates the movement of batsmen between wickets.\\
        (RAM: 1--8) & 2. Crease and stump movement: Regulates player positioning near the crease and stumps.\\
        (RPM: 1--5) & 3. Fielding position restrictions: Specifies fielding positions and limitations during play.\\
        & 4. Fielding the ball: Regulates the legal method of fielding and returning the ball.\\
        & 5. Running in the protected area: Prohibits running in areas designated for protection.\\
        & 6. Backfoot no-ball rule: Penalizes bowlers for overstepping the crease during delivery.\\
        & 7. Fair and unfair play: Governs fair play conduct and penalizes unfair actions on the field.\\
        & 8. Pitch etiquette: Specifies behavior and conduct on the pitch during play.\\
    \hline
    Field Hockey 
        & 1. Offside: Regulates player positioning relative to the opponents during play.\\
        (RAM: 1--9) & 2. Advantage: Allows play to continue despite a foul, benefiting the fouled team.\\
        (RPM: 1--7) & 3. Obstruction: Prohibits players from blocking opponents' access to the ball.\\
        & 4. Dangerous play: Penalizes players for actions that may endanger themselves or others.\\
        & 5. Pushing: Regulates the legal method of using the stick to push the ball.\\
        & 6. Impeding: Prohibits players from obstructing opponents' movement without the ball.\\
        & 7. Illegal tackling: Penalizes players for improper tackles (e.g., with excessive force).\\
        & 8. Back stick: Penalizes players for using the rounded backside of the stick to play the ball.\\
        & 9. High sticks: Prohibits players from playing the ball above shoulder height with their stick.\\
    \hline
    Futsal  
        & 1. Running with the ball: Regulates dribbling and movement with the ball.\\
        (RAM: 1--8) & 2. 3-Second rule: Limits the time a player can hold the ball without dribbling or passing.\\
        (RPM: 1--5) & 3. Goalkeeper restrictions: Specifies rules and limitations unique to the goalkeeper position.\\
        & 4. Encroachment on free kicks: Prohibits players from encroaching during free kicks.\\
        & 5. Kicking in: Governs the method of restarting play from the touchline.\\
        & 6. Intentional time wasting: Penalizes teams for deliberately wasting time during play.\\
        & 7. Five-foul limit: Penalizes teams for committing a certain number of fouls in a half.\\
        & 8. No slide tackling: Prohibits slide tackling to ensure player safety.\\
    \hline
    Handball 
        & 1. Dribbling: Regulates the movement of the ball while in hand possession.\\
        (RAM: 1--8) & 2. Passive play: Prohibits stalling or delaying the game without scoring attempts.\\
        (RPM: 1--7) & 3. Stepping inside the goal area: Limits players from entering the goal area during play.\\
        & 4. Jumping: Regulates jumping actions, particularly during shooting or passing.\\
        & 5. Encroachment on free throws: Restricts opponents from entering the free-throw area.\\
        & 6. Goalkeeper restrictions: Specifies rules and limitations for the goalkeeper position.\\
        & 7. Holding and pushing: Penalizes players for holding or pushing opponents illegally.\\
        & 8. Illegal screening: Prohibits players from obstructing defenders during set plays.\\
    \hline
    Ice Hockey  
    & 1. Offside: Prevents attacking players from entering the offensive zone before the puck.\\
    (RAM: 1--9) & 2. Interference: Prohibits obstructing or impeding an opponent not in possession of the puck.\\
    (RPM: 1--6) & 3. Hooking: Restricts players from using their stick to hook an opponent.\\
    & 4. Slashing: Penalizes players for swinging their stick at an opponent.\\
    & 5. Boarding: Penalizes players for checking an opponent into the boards violently.\\
    & 6. Charging: Penalizes players for charging into an opponent violently.\\
    & 7. Icing: Regulates the clearing of the puck from one end of the rink to the other.\\
    & 8. Tripping: Penalizes players for causing opponents to fall by tripping them.\\
    & 9. High-sticking: Bars players from using sticks above shoulder height.\\    
    \hline
    Lacrosse 
    & 1. Offside: Regulates player positioning on the field during play.\\
    (RAM: 1--8) & 2. Moving picks or screens: Prohibits illegal screens to obstruct defenders.\\
    (RPM: 1--7) & 3. Crease violation: Penalizes players for entering the crease area during play.\\
    & 4. Stick checks: Regulates legal stick checking actions during play.\\
    & 5. Offensive fouls: Penalizes offensive players for illegal actions during play.\\
    & 6. Illegal picks: Prohibits illegal screens set by offensive players.\\
    & 7. Interference: Penalizes players for obstructing opponents without playing the ball.\\
    & 8. Illegal body checking: Prohibits excessive or dangerous body checking.\\
    \hline
    Roller Hockey 
    & 1. Offside: Regulates player positioning relative to the puck during play.\\
    (RAM: 1--8) & 2. Obstruction: Prohibits obstructing opponents without playing the puck.\\
    (RPM: 1--4) & 3. Crease violation: Penalizes players for entering the crease area during play.\\
    & 4. Tripping: Penalizes players for causing opponents to fall by tripping them.\\
    & 5. Illegal contact: Penalizes players for illegal physical contact with opponents.\\
    & 6. Illegal screen: Prohibits illegal screens to obstruct opponents.\\
    & 7. Delay of game: Penalizes teams for delaying the game intentionally.\\
    & 8. High-sticking: Prohibits contacting the puck with the stick above shoulder height.\\
    \hline
    Rugby  
    & 1. Offside: Regulates player positioning during set plays and general play.\\
    (RAM: 1--12) & 2. Advantage: Allows play to continue for minor infractions.\\
    (RPM: 1--7) & 3. Maul: Governs the formation and legality of mauls during play.\\
    & 4. Scrum: Regulates the engagement and conduct of scrums to restart play.\\
    & 5. Lineout: Specifies rules and procedures for lineouts to restart play from touch.\\
    & 6. Kicking: Regulates kicking actions during open play and set pieces.\\
    & 7. Not retreating 10 meters: Penalizes teams for insufficient retreat from penalties.\\
    & 8. Tackling: Governs legal tackling techniques and player safety during tackles.\\
    & 9. Ruck: Regulates player actions and roles in rucks formed during play.\\
    & 10. Illegal blocking or holding: Prohibits illegal blocking or holding actions during play.\\
    & 11. Dangerous play: Penalizes players for actions that endanger opponents' safety.\\
    & 12. Not binding correctly: Regulates the correct binding of players in scrums and mauls.\\
    \hline
    Soccer 
    & 1. Offside: Prevents forward players from receiving the ball behind the defense.\\
    (RAM: 1--9) & 2. Handling: Prohibits using hands or arms to control the ball, except for the goalkeeper.\\
    (RPM: 1--7) & 3. Goalkeeper: Specifies actions and limitations unique to the goalkeeper position.\\
    & 4. Impeding: Restricts obstructing an opponent's movement without playing the ball.\\
    & 5. Fouls against goalkeepers: Protects goalkeepers from physical contact.\\
    & 6. Charging opponents: Prohibits excessive or dangerous body contact with opponents.\\
    & 7. Impeding opponents: Prevents deliberately impeding an opponent's progress on the field.\\
    & 8. Fouls and misconduct: Governs various rule violations, including fouls and misconduct.\\
    & 9. Simulation and diving: Penalizes players for simulating fouls or exaggerating contact.\\
    \hline
    Volleyball 
    & 1. Rotational faults: Regulates player positions during serve and rotation.\\
    (RAM: 1--8) & 2. Foot faults during service: Prevents foot faults during the serving motion. \\
    (RPM: 1--7) & 3. Back-row attack violations: Limits back-row players from attacking past the 3-meter line.\\
    & 4. Blocking across the net: Governs legal blocking actions across the net.\\
    & 5. Center line violations: Prohibits players from crossing the center line during play.\\
    & 6. Libero restrictions: Specifies actions and limitations for the libero player.\\
    & 7. Illegal substitutions: Regulates player substitutions and entry onto the court.\\
    & 8. Net violations: Penalizes players for touching the net during play. \\
    \hline
    Water Polo  
    & 1. Holding or pushing: Prohibits holding or pushing opponents in the water.\\
    (RAM: 1--7) & 2. Interference: Penalizes players for obstructing opponents without playing the ball.\\
    (RPM: 1--5) & 3. Sinking: Prohibits players from deliberately sinking or diving to gain advantage.\\
    & 4. Kick-off: Specifies rules and procedures for the kick-off to start the game.\\
    & 5. Corner throw: Governs the method of restarting play from the corner of the pool.\\
    & 6. Striking: Penalizes players for striking opponents with hands or arms.\\
    & 7. Two-handed pushoff: Regulates the legal use of hands to push off opponents.\\
    \hline
\end{longtable}
}

\end{document}